\documentclass[a4paper,11pt]{article}
\pdfoutput=1 

\usepackage{jheppub} 
\usepackage{bm}
\usepackage{cancel}
\usepackage[T1]{fontenc} 
\usepackage{slashed}
\newcommand{\ri}{{\rm i}}

\title{\boldmath Extra-natural production of superheavy Kaluza-Klein particles}

\author[a]{Yusuke Yamada}

\affiliation[a]{Waseda Institute for Advanced Study, Waseda University, 1-21-1 Nishi Waseda, Shinjuku, Tokyo 169-0051, Japan}
\affiliation[b]{Cosmology, Gravity, and Astroparticle Physics Group, Center for Theoretical Physics of the Universe,
Institute for Basic Science (IBS), Daejeon, 34126, Korea}
\emailAdd{y-yamada@aoni.waseda.jp}

\abstract{Gauge fields in extra compact dimensions can drive inflation in the four-dimensional (4D) non-compact spacetime, a scenario known as {\it extra-natural inflation}. A time-dependent gauge field configuration generates the electric field along the compact dimension, enabling the production of Kaluza–Klein (KK) particles charged under the field via the Schwinger effect. We construct the extra-natural inflation model within a five-dimensional (5D) quantum electrodynamics (QED) framework coupled to gravity including matter fields that generate the inflationary one-loop effective potential. In general, multiple charged fields can exist, and we show that KK particle production occurs under these conditions. Since KK momentum is conserved, the produced KK particles may become superheavy dark matter or dominate the universe, depending on the model parameters. Furthermore, we show that even when the gauge field acts not as the inflaton but as a spectator field, its post-inflationary oscillations, initiated when the Hubble friction becomes negligible, can also generate superheavy KK modes. This suggests that KK particle production is a generic outcome when gauge potentials along compact dimensions are light. }

\begin{document} 
\maketitle
\flushbottom
\section{Introduction}

The existence of extra spaces is one of the most important predictions of superstring theory, a leading candidate for a quantum theory of gravity. Collider experiments and cosmological observations have not detected any signatures of extra spatial dimensions, and therefore, such dimensions are thought to be compactified with small radii to remain consistent with current experimental constraints. If compact spaces are small enough, modes associated with compact dimensions—called Kaluza-Klein modes (KK modes)—acquire four-dimensional masses of the order of the inverse compact space radii. Such modes cannot be excited below the compactification scale and can be integrated out, yielding a 4D effective theory.

The gauge fields along compact dimensions may have a relatively flat potential in the 4D effective theory because gauge symmetry forbids tree-level potential terms. Such fields may play the role of the inflaton, driving accelerated expansion of the three (non-compact) spatial dimensions in the early universe~\cite{Guth:1980zm,Starobinsky:1980te,Sato:1981qmu,Linde:1981mu,Albrecht:1982wi}, a scenario known as “extra-natural inflation”~\cite{Arkani-Hamed:2003xts,Arkani-Hamed:2003wrq,Kaplan:2003aj}. The original 5D model on $\mathbb R^{1,3}\times S^1$ is based on a U(1) gauge theory, and the gauge field along $S^1$ is identified as the inflaton. The inflaton potential originates from the one-loop effective potential induced by particles charged under the U(1) symmetry, making it dependent on the matter content of the theory.

In our recent work~\cite{Yamada:2024aca,Abe:2024mka},\footnote{See also \cite{Furuuchi:2015foh} for a related work.} we pointed out that the time-dependent gauge potential configurations realized in extra-natural inflation lead to “electric fields” along compact spaces, and KK momenta can be decelerated and accelerated along compact spaces, leading to the production of KK particles from the vacuum, a process we call the KK-Schwinger effect. The mechanism is essentially the same as the 4D Sauter-Schwinger effect~\cite{Sauter:1931zz,Schwinger:1951nm} in strong field QED except that the KK momenta are quantized due to the periodicity of the compact spaces. For example, in a constant electric field along the compact space, the $n$-th KK momentum of a charged field at time $t$ is given by
$$P_n^2=M_{KK,n}^2=\frac{1}{R^2}(n-gE_y Rt)^2 $$
where $g$ is the gauge coupling, $E_y$ the electric field along the compact direction, and $R$ the compactification radius. Physically, the mode undergoes acceleration or deceleration due to the electric field. In general, there is a time when the KK momentum vanishes. At that moment, the energy of the KK mode significantly decreases, and the mode can be produced from vacuum non-perturbatively. The most important property of the KK-Schwinger effect is that the production rate is independent of the KK mass scale because the KK momentum vanishes at production. As the gauge potential evolves, the produced particles become heavier. Therefore, even if the Hubble parameter during and after inflation is much smaller than the compactification scale, superheavy KK particles can still be produced due to the KK-Schwinger effect. Note that the KK momentum is a conserved quantity as long as the compact space possesses translational invariance, and therefore, KK particles cannot decay into zero modes however heavy they are. They may therefore serve as superheavy dark matter candidates.

In this work, we construct a concrete inflationary model realizing both radion stabilization and inflation within a simple 5D QED framework coupled to gravity. Our model remains a toy model from the ultraviolet (UV) completion perspective, yet has some advantages. First, our field theory model is fully calculable and can be generalized to higher-dimensional models. We treat the inflaton and gravitational fields as classical, while scalar and spinor fields are quantized; their expectation values determine the effective potential for the classical fields. Although we do not address the backreaction effects of matter quantum fields on the classical fields in this paper, our formulation can, in principle, accommodate such semi-classical dynamics.

We find that multiple charged fields are required to “tune” the potential, leading to multi-natural inflation~\cite{Czerny:2014wza,Czerny:2014xja,Higaki:2014pja,Higaki:2014mwa}, capable of producing an effective potential yielding the spectral index $n_s$ and the tensor-to-scalar ratio $r$ consistent with recent cosmic microwave background (CMB) observations. We further show that introducing multiple charged fields generally leads to the KK-Schwinger effect, whereas a model with a single charged sector, producing a purely sinusoidal effective potential, is unlikely to do so.

Depending on model parameters, a significant abundance of superheavy KK particles may be produced, potentially explaining present-day dark matter or even dominating the universe, posing potential challenges for inflationary models in higher-dimensional spacetime theories. Furthermore, even when the gauge field is not the inflaton but a light spectator with a non-zero expectation value during inflation, its oscillations—beginning when the Hubble parameter falls to its mass scale—can trigger superheavy KK particle production. This suggests that a light Wilson line field in the early universe can produce such particles if its motion (i.e., the induced electric field) changes the KK momentum over very long timescales.

Before closing the introduction, we comment on connections with other mechanisms for superheavy particle production during and after inflation. Due to unitarity, thermal processes cannot produce sufficient dark matter heavier than about $300$ TeV~\cite{Griest:1989wd}. However, non-thermal production is still possible. Cosmological particle creation effects~\cite{Parker:1969au,Parker:1971pt} provide a mechanism for producing heavy particles~\cite{Kuzmin:1998uv,Chung:1998zb,Chung:1998ua,Kolb:1998ki,Chung:2001cb,Ling:2025nlw}, where the rapid expansion of the universe generates particles from the vacuum non-perturbatively via gravitational effects.\footnote{See \cite{Kolb:2023ydq} for a recent review of cosmological particle production and its applications. See also the reprinted version of the Ph.D. thesis of Parker~\cite{Parker:2025jef} as a reference on cosmological particle production.} A similar but different mechanism to cosmological particle production is preheating after inflation, where oscillations of the inflaton field non-perturbatively produce coupled quantum particles. Furthermore, particles produced during preheating can be much heavier than the Hubble scale during inflation, and their decay at the time of maximal mass leads to instant preheating~\cite{Felder:1998vq,Felder:1999pv}.\footnote{Quantum field theoretical treatments of instant preheating have been recently discussed in \cite{Taya:2022gzp}.} Among these mechanisms, the KK-Schwinger effect is similar to preheating, but the masses of the produced particles end up large, whereas in standard preheating the effective masses decay as the inflaton oscillation amplitude decreases.

The rest of the paper is organized as follows. In Sec.\ref{Setup}, we show the Lagrangian of the 5D QED system under consideration. We KK-expand 5D fields and then quantize spin 0 and 1/2 fields. Then, we discuss extra-natural inflation within a quantum field theory framework coupled to background gravitational and electromagnetic fields. We first show the set of dynamical equations and explain that a multi-charge extension is necessary to deform the potential such that inflationary observables are compatible with recent CMB data. Then, we construct an illustrative example in Sec.\ref{illustrative} and show that the KK-Schwinger effect produces KK particles during and after inflation in Sec.\ref{KKSchwinger}. We briefly discuss how the produced KK particles affect the universe after inflation. In particular, we derive a constraint on the produced KK particle energy density in Sec.\ref{fateofKK}. We then show another scenario where the Wilson line field is not an inflaton but a spectator field, and show that the KK particles are produced from the vacuum in the same way as in extra-natural inflation. Finally, we conclude in Sec.\ref{conclusion}. In Appendix\ref{notation}, we present our spinor conventions. We evaluate the energy momentum tensor operators in Appendix~\ref{EMtensor}, used in the main text. We also show the complete form of the equations of motion without any approximation in Appendix~\ref{completeEOM}, including the Hubble-induced corrections as well as backreaction terms of quantum fields. We briefly comment on the gradient energy or the four-form flux potential, as an alternative to the 5D cosmological constant for realizing a positive cosmological constant in the 4D effective theory, in Appendix~\ref{gradE}. In Appendix~\ref{illustative2}, we give a set of model parameters for extra-natural inflation different from the one in Sec.\ref{illustrative}, where the KK particle energy produced after inflation is small enough to avoid the constraint given in Sec.\ref{fateofKK}.
\section{Setup: 5D QED model coupling to gravity}\label{Setup}
We consider a 5D QED model in the time-dependent cosmological background where the action is
\begin{align}
    S_{\rm tot}=&\int d^5x\sqrt{-G}\Biggl[\frac{M_5^3}{2}{\cal R}_5-\Lambda^5+\sum_A\left(-D_M\Phi^AD^M
    \Phi^{A\dagger}-m_{0,A}^2|\Phi^A|^2\right)\nonumber\\
  &\qquad +\sum_B\left(-\overline{\Psi}{}^B\Gamma^M D_M\Psi^B+m_{1/2,B}\overline{\Psi}{}^B\Psi^B\right) -\frac14F_{MN}F^{MN} \Biggr],
\end{align}
where ${\cal R}_5$ denotes the 5D Ricci scalar, $M_5$ the 5D Planck scale, $\Lambda^5$ the 5D cosmological constant, $\Phi^A$ and $\Psi^B$ the scalar and the spinor fields with respective labels $A,B$, and $m_{0,A}$ and $m_{1/2,B}$ are scalar and spinor bulk masses. The last term is the U(1) gauge field action, and $D_M$ denotes the covariant derivative under both diffeomorphism and the U(1) gauge symmetry.
We will consider $S^1$ compactification of 5D spacetime with a metric given by
\begin{align}
    ds^2=G_{MN}dx^Mdx^N=\frac{1}{b(t)}\left(-dt^2+a^2(t)d\bm x^2\right)+b^2(t)dy^2,
\end{align}
with the periodic condition $y\sim y+2\pi R$ where $R$ is a radius parameter. The above metric choice leads to a constant 4D Planck scale: Substituting the above metric into the Einstein-Hilbert action reads
\begin{align}
    S_{\rm EH}=&\frac{M_5^3}{2}\int d^4x dy\sqrt{-G}{\cal R}_5=\frac{2\pi RM_5^3}{2}\int dt d^3xa^3\left[6H_a^2+\frac32H_b^2\right],\label{EHaction}
\end{align}
where $H_a\equiv \frac{\dot a}{a}$ and $H_b\equiv \frac{\dot b}{b}$ , the dot denotes the time derivative $\partial_t$ and we have omitted the total time-derivative terms. Then, we may identify the 4D Planck scale as $M_{\rm pl}^2=2\pi R M_5^3$.\footnote{This does not mean the compact space has a fixed volume as it is given by $L_{\rm phys}=\int \sqrt{G_{55}}dy=2\pi Rb$. Instead, the change of the physical volume of the compact direction is absorbed by the conformal rescaling of 4D spacetime within our choice, which makes the 4D Planck scale constant.} We parametrize the extra-space scale factor as $b=\exp\left(\sqrt{\frac23}\varphi/M_{\rm pl}\right)$, which reads
\begin{align}
    S=\int dt d^3xa^3 \left[3M_{\rm pl}^2H_a^2+\frac12 \dot \varphi^2\right]
\end{align}

We also consider a background U(1) gauge field $A_5(t)$. Although we will discuss the canonical normalization of a spin 1 field fluctuation, we first make the background zero mode a canonical 4D real scalar, which can be achieved as follows: The quadratic action is formally given by
\begin{align}
   \int d^5x \sqrt{-G}\left[-\frac14 F_{MN}F^{MN}\right]\supset \int d^4xdy\frac{a^3}{2b}\left(\frac{1}{b}\dot{A}_5^2\right)=\int d^4xa^3\left(\frac{2\pi R}{2b^2}\dot{A}_5^2\right).
\end{align}
Then, it is reasonable to define
\begin{align}
    \vartheta\equiv \frac{\sqrt{2\pi R}}{b}A_5,
\end{align}
and the above action becomes
\begin{align}
    \int d^4xa^3\frac12\left(\dot\vartheta+{H}_b\vartheta\right)^2
    =\int d^4xa^3\frac12\left[\dot\vartheta^2-\left(\dot{H}_b+3H_aH_b-H_b^2\right)\vartheta^2\right],
\end{align}
up to total time derivative terms.
The new field $\vartheta$ is yet non-canonical in the usual sense, but is canonical when the radion $b$ is stabilized. The covariant derivative of a charged matter is 
\begin{align}
    D_5=\partial_5-\ri qgA_5+\cdots,
\end{align}
where $g$ is a coupling constant with its mass dimension $-1/2$, and the ellipses denote other gauge connections. For later convenience, we define a dimensionless gauge coupling constant as
\begin{align}
    g=\sqrt{2\pi R} g_0,
\end{align}
which makes the combination $gA_5=g_0b\vartheta$. 

Let us introduce charged matter fields of spin $0$ and $1/2$ coupling to the background homogeneous fields $b(t)$ and $\vartheta(t)$. We first consider a complex scalar field in this background,
\begin{align}
   S_0=& \int d^4x dy \sqrt{-G}\left[-D_M\Phi g^{MN}D_N\Phi^\dagger-m_0^2\Phi\Phi^\dagger\right]\nonumber\\
   =&\int d^4x dy a^3\left[|\dot\Phi|^2-\frac{1}{a^2}|\partial_i\Phi|^2-\frac{1}{b^3}|D_y\Phi|^2-\frac{m_0^2}{b}|\Phi|^2\right],\label{5DPhiaction}
\end{align}
where $D_M\Phi=\left(\partial_M-\ri qgA_M\right)\Phi$, $q$ is the U(1) charge of $\Phi$, $g$ is a gauge coupling, and we have assumed that only $A_5$ has a nontrivial background value. We KK-expand the complex scalar field $\Phi$ and $\Phi^\dagger$ as
\begin{align}
    \Phi(x^\mu,y)=&\frac{1}{\sqrt{2\pi R}}\sum_{n\in \mathbb Z}\tilde\phi_n(x)e^{+\frac{\ri n y}{R}},\label{0KKexpansion}\\
    \Phi^\dagger(x^\mu,y)=&\frac{1}{\sqrt{2\pi R}}\sum_{n\in \mathbb Z}\tilde\phi^\dagger_n(x)e^{-\frac{\ri n y}{R}}.
\end{align}
By substituting it into \eqref{5DPhiaction}, we find
\begin{align}
   S_0=&\int d^4x a^3\sum_{n\in \mathbb Z}\left[\left|\dot{\tilde\phi}_n\right|^2-\frac{1}{a^2}|\partial_i\tilde{\phi}_n|^2-\frac{1}{b^3}\left|\left(\frac{n}{R}-qg_0b\vartheta\right)\tilde{\phi}_n\right|^2-\frac{m_0^2}{b}|\tilde{\phi}_n|^2\right]\nonumber\\
  =&\int d^4x \sum_{n\in \mathbb Z}\left[\left|\dot{\phi}_n-\frac32H_a\phi_n\right|^2-\frac{1}{a^2}|\partial_i\phi_n|^2-\frac{1}{b^3}\left|\left(\frac{n}{R}-qg_0b\vartheta\right)\phi_n\right|^2-\frac{m_0^2}{b}|\phi_n|^2\right]\nonumber\\
   =&\int d^4x \sum_{n\in \mathbb Z}\left[\left|\dot{\phi}_n\right|^2-|\partial_i\phi_n|^2-M^2_{0,n}|\phi_n|^2\right]\label{Sscalar}
\end{align}
where we have defined $\phi_n\equiv a^{\frac32}\tilde\phi_n$ and 
\begin{align}
    M_{0,n}^2\equiv&M_{KK,n}^2+\frac{m_0^2}{b}-\frac94H_a^2-\frac32\dot{H}_a,\\
    M_{KK,n}^2\equiv&\frac{1}{b^3}\left(\frac{n}{R}-qg_0b\vartheta\right)^2.\label{MKK0}
\end{align}
Thus, we have obtained the canonical KK decomposition in the time-dependent background.

Similarly, the 5D spinor field action can be written as
\begin{align}
    S_{1/2}=&\int d^4x dy\sqrt{-G}\left[-\bar\Psi\gamma^A e^M_{A}D_M\Psi+m_{1/2}\bar\Psi\Psi\right],
\end{align}
where $A=\mathtt0,\mathtt1,\mathtt2,\mathtt3,\mathtt5$ denotes the local Lorentz index,\footnote{In the following, we will denote the local Lorentz indices by the typewriter style numbers or letters, and the curved ones are denoted in the normal style.} the vielbein is
\begin{align}
    e^M_A=
{\rm diag}\left(b^{\frac12}(t),\frac{b^{\frac12}(t)}{a(t)},\frac{b^{\frac12}(t)}{a(t)},\frac{b^{\frac12}(t)}{a(t)},\frac{1}{b(t)}\right),
\end{align}
and the covariant derivative is
\begin{align}
    D_M\Psi=\left(\partial_M+\frac{1}{4}\omega_M^{AB}\gamma_{AB}-\ri qA_M\right)\Psi,
\end{align}
with $\gamma_{AB}\equiv\frac12(\gamma_A\gamma_B-\gamma_B\gamma_A)$ and the non-vanishing spin connections
\begin{align}
    \omega^{\mathtt k\mathtt 0}_{\mathtt l}=-\left(H_a-\frac12H_b\right)b^{\frac12}\delta_{\mathtt l}^{\mathtt k},\quad \omega^{\mathtt 5\mathtt0}_{\mathtt 5}=-H_bb^{\frac12},
\end{align}
where  $\mathtt k,\mathtt l=\mathtt1,\mathtt2,\mathtt3,\mathtt5$ denotes the local flat index. Then, we explicitly obtain
\begin{align}
    \gamma^A e_A^MD_M\Psi
    =&b^{\frac12}\left(\gamma^{\mathtt0}\dot\Psi+\frac{1}{a}\gamma^{\mathtt i}\partial_{i}\Psi+\frac{1}{2}\left(3H_a-\frac12H_b\right)\gamma^{\mathtt0}\Psi+b^{-\frac32}\gamma^{\mathtt 5}\left(\partial_y-\ri q A_5\right)\Psi\right)\nonumber\\
=& a^{-\frac52}b^{\frac34}\left(\gamma^{\mathtt 0}\dot\Psi_c+\frac1a\gamma^{\mathtt i}\partial_{i}\Psi_c+b^{-\frac32}\gamma^{\mathtt 5}\left(\partial_y-\ri qg A_5\right)\Psi_c\right),
\end{align}
where $\Psi=a^{-\frac32}b^{+\frac14}\Psi_c$.
Thus, the Lagrangian is rewritten as
\begin{align}
    S_{1/2}=&\int d^4x dy\sqrt{-G}\left[-\bar\Psi\gamma^A e^M_{A}D_M\Psi+m_{1/2}\bar\Psi\Psi\right]\nonumber\\
    =&\int d^4x dy\left[-\bar\Psi_c\left(\gamma^{\mathtt 0}\dot{\Psi}_c+\frac1a\gamma^{\mathtt i}\partial_{i}\Psi_c+\frac{1}{b^\frac32}\gamma^{\mathtt 5}\left(\partial_y-\ri q gA_5\right)\Psi_c\right)+\frac{m_{1/2}}{b^{\frac12}}\bar\Psi_c\Psi_c\right].
\end{align}
We expand the spinor field as
\begin{align}
    \Psi_c(x,y)=\frac{1}{\sqrt{2\pi R}}\sum_{n\in \mathbb Z}\psi_n(x)e^{\frac{\ri ny}{R}}.
\end{align}
Substituting the KK expansion, we obtain
\begin{align}
    S_{1/2}=\int d^4x \sum_{n\in\mathbb Z}\left[-\bar\psi_n\left(\gamma^{\mathtt 0}\dot\psi_n+\frac1a\gamma^{\mathtt i}\partial_{i}\psi_n+\frac{\ri }{b^{\frac32}} \gamma^{\mathtt 5}\left(\frac{n}{R}- qg_0b\vartheta\right)\psi_n\right)+\frac{m_{1/2}}{b^{\frac12}}\bar\psi_n\psi_n\right].
\end{align}
 We further decompose the 4D Dirac spinors $\psi_n$ by two-component ones as
\begin{align}
    \psi_n=\left(\begin{array}{c}\chi_{n\alpha}\\ \zeta^{\dagger\dot\alpha}_n\end{array}\right),
\end{align}
where $\alpha,\dot\alpha$ are two-component spinor indices. Note that the Dirac conjugate spinors are
\begin{align}
    \bar{\psi}_n=(\zeta^\alpha_n,{\chi}^\dagger_{n\dot\alpha})=\ri\psi_n^{\dagger}\gamma^{\mathtt 0}.
\end{align}
Then, the action can be rewritten as
\begin{align}
    S_{1/2}=&\int d^4x \sum_{n\in\mathbb Z}\Biggl[-\ri\zeta_n\sigma^0\partial_0{\zeta}^\dagger_n-\ri\chi_{n}\sigma^0\partial_0\chi^\dagger_n-\frac{\ri}{a}\zeta_n\sigma^i\partial_i{\zeta}_n^\dagger-\frac{\ri}{a}\chi_{n}\sigma^i\partial_i\chi_n^\dagger\nonumber\\
   &\hspace{2cm} + M_{1/2,n}(t)\zeta_n\chi_n+M^*_{1/2,n}(t){\zeta}^\dagger_n{\chi}^\dagger_n \Biggr]\label{Sspinor}
\end{align}
where
\begin{align}
  M_{1/2,n}=& \frac{m_{1/2}}{b^{\frac12}}-\frac{\ri }{b^{\frac32}}\left(\frac{n}{R}-qg_0b\vartheta\right).
\end{align}
The spinor fields are not affected by the expansion of the universe, and hence the Hubble-induced mass terms do not appear in the fermion mass.

\subsection{Quantization}
\subsubsection{scalar fields}
For scalar fields, we expand quantum fields as
\begin{align}
  \hat{\phi}_n(t,\bm x) =&\int \frac{d^3\bm k}{(2\pi)^{\frac32}}\left[\hat{a}_{n,\bm k}f_{n,k}(t)e^{\ri \bm k\cdot\bm x}+\hat{b}_{n,\bm k}^\dagger f^*_{n,k}(t)e^{-\ri \bm k\cdot \bm x}\right],\nonumber\\
  \hat{\phi}_n^{\dagger} (t,\bm x)=&\int \frac{d^3\bm k}{(2\pi)^{\frac32}}\left[\hat{b}_{n,\bm k}f_{n,k}(t)e^{\ri \bm k\cdot\bm x}+\hat{a}_{n,\bm k}^\dagger f^*_{n,k}(t)e^{-\ri \bm k\cdot \bm x}\right],\label{spin0Q}
\end{align}
where the set of the creation and annihilation operators $(\hat{a}^\dagger_{n,\bm k}, \hat{a}_{n,\bm k})$ and $(\hat{b}^\dagger_{n,\bm k}, \hat{b}_{n,\bm k})$ satisfy the commutation relation $[\hat{a}_{n,\bm k},\hat{a}^\dagger_{n',\bm k'}]=\delta_{nn'}\delta^3(\bm k-\bm k')=[\hat{b}_{n,\bm k},\hat{b}^\dagger_{n',\bm k'}]$, the mode function $f_{n,k}(t)$ satisfies
\begin{align}
    \ddot{f}_{n,k}(t)+\omega_{0,n,k}^2(t)f_{n,k}(t)=0,
\end{align}
where $|\bm k|=k$ and 
\begin{align}
\omega_{0,n,k}^2(t)=\frac{k^2}{a^2(t)}+M_{0,n}^2(t).
\end{align}
We write the formal solution to the mode equation as
\begin{align}
    f_{n,k}(t)=\frac{\alpha_{n,k}(t)}{\sqrt{2\omega_{0,n,k}(t)}}e^{-\ri\int^t dt'\omega_{0,n,k}(t')}+\frac{\beta_{n,k}(t)}{\sqrt{2\omega_{0,n,k}(t)}}e^{+\ri\int^t dt'\omega_{0,n,k}(t')},
\end{align}
with the auxiliary functions $\alpha_{n,k}(t)$ and $\beta_{n,k}(t)$, which satisfy the normalization condition
\begin{align}
    |\alpha_{n,k}(t)|^2-|\beta_{n,k}(t)|^2=1.
\end{align}
One can confirm that the formal solution is consistent with the canonical commutation relation $[\hat\phi_n(t,\bm x),\hat\pi_{n,\phi}(t,\bm y)]=[\hat{\phi}_{n}(t,\bm x),\dot{\hat{\phi}}{}^\dagger_{n}(t,\bm y)]=\ri\delta^3(\bm x-\bm y)$. For the consistency with the mode equations, the auxiliary functions satisfy
\begin{align}
 \frac{d}{dt} \left(\begin{array}{c}
        \alpha_{n,k}\\ \beta_{n,k}
   \end{array}\right)=\left(\begin{array}{cc}
      0 & \frac{\dot{\omega}_{0,n,k}}{2\omega_{0,n,k}}e^{+2\ri\int^t dt'\omega_{0,n,k}(t')} \\
     \frac{\dot{\omega}_{0,n,k}}{2\omega_{0,n,k}}e^{-2\ri\int^t dt'\omega_{0,n,k}(t')}   & 0
   \end{array}\right)\left(\begin{array}{c}
       \alpha_{n,k}\\ \beta_{n,k}
   \end{array}\right)
\end{align}
However, the presence of the phase factor on the right-hand side is inconvenient for practical calculations, and we define $\tilde{\alpha}_{n,k}\equiv \alpha_{n,k}e^{-\ri\int^t dt'\omega_{0,n,k}(t')}$ and $\tilde{\beta}_{n,k}\equiv \beta_{n,k}e^{+\ri\int^t dt'\omega_{0,n,k}(t')}$, with which the above equation becomes
\begin{align}
  \frac{d}{dt} \left(\begin{array}{c}
        \tilde{\alpha}_{n,k}\\ \tilde{\beta}_{n,k}
   \end{array}\right)=\left(\begin{array}{cc}
      \ri\omega_{0,n,k}  & \frac{\dot{\omega}_{0,n,k}}{2\omega_{0,n,k}} \\
     \frac{\dot{\omega}_{0,n,k}}{2\omega_{0,n,k}}   & -\ri\omega_{0,n,k}
   \end{array}\right)\left(\begin{array}{c}
        \tilde{\alpha}_{n,k}\\ \tilde{\beta}_{n,k}
   \end{array}\right).\label{tildeab}
\end{align}
We emphasize that there is no unique way to define the instantaneous vacuum mode functions as a consequence of the lack of time-translation invariance. As basis functions, we may choose other WKB functions of the higher adiabatic order. The choice of the adiabatic basis changes the behavior of the particle number at the intermediate time when the time-dependence of the backgrounds is non-vanishing~\cite{Dabrowski:2014ica,Dabrowski:2016tsx,Yamada:2021kqw}.

\subsubsection{spinor fields}
We quantize spinor fields that have time-dependent one-particle energy. We formally decompose the spinor field operators as
\begin{align}
    \hat{\chi}_n^\alpha(t,\bm x)=&\sum_{h=\pm}\int \frac{d^3\bm k}{(2\pi)^\frac32}\left[\hat{c}_{n,\bm k,h}\eta_{n,k,h}(t)e^{\ri\bm k\cdot\bm x}\xi_h^\alpha(\hat{\bm k})+\hat{d}^\dagger_{n,\bm k,h}\bar{\lambda}_{n,k,h}(t)e^{-\ri{\bm k}\cdot\bm x}\xi_{h\dot\beta}^{\dagger}(\hat{\bm k})\bar{\sigma}_{\mathtt0}^{\dot\beta\alpha}\right],\label{chi}\\
    \hat{\zeta}_n^\alpha(t,\bm x)=&\sum_{h=\pm}\int \frac{d^3\bm k}{(2\pi)^\frac32}\left[\hat{d}_{n,\bm k,h}\eta_{n,k,h}(t)e^{\ri\bm k\cdot\bm x}\xi_h^\alpha(\hat{\bm k})+\hat{c}^\dagger_{n,\bm k,h}\bar{\lambda}_{n,k,h}(t)e^{-\ri{\bm k}\cdot\bm x}\xi_{h\dot\beta}^{\dagger}(\hat{\bm k})\bar{\sigma}_{\mathtt0}^{\dot\beta\alpha}\right]\label{zeta},
\end{align}
where we have introduced the helicity eigenspinors $\xi_h$ satisfying
\begin{align}
  & ((\hat{\bm k}\cdot{\bm\sigma}){\bm\xi}_h)_{\alpha}=\hat{k}_{\mathtt i}(\sigma^{\mathtt0}_{\alpha\dot\alpha}(\bar{\sigma}^{\mathtt i})^{\dot\alpha\beta})\xi_\beta^{h}(\hat{\bm k})=h\xi^{h}_\alpha(\hat{\bm k}),\\
 & ({\bm \xi}_h^\dagger(\hat{\bm k}\cdot{\bm\sigma}))_{\dot\alpha} =\xi^\dagger_{h\dot\beta}(\hat{\bm k})(\bar{\sigma}_{\mathtt0})^{\dot\beta\alpha}(\hat{k}_{\mathtt i}\sigma^{\mathtt i})_{\alpha\dot\alpha}=h\xi_{h\dot\alpha}^{\dagger}(\hat{\bm k}),\\
&\xi_{h\dot\beta}^{\dagger}(-\hat{\bm k})\bar{\sigma}_{\mathtt0}^{\dot\beta\alpha}\equiv \iota_h(\hat{\bm k})\xi_{h}^\alpha(\hat{\bm k}),\label{flip}
\end{align}
where $h=\pm1$ is the helicity, $\hat{k}_i$ is a unit three-momentum vector satisfying $|\hat{\bm k}|^2=1$, and $\iota_h(\hat{\bm k})$ is a phase factor satisfying $\iota_h(-\hat{\bm k})=-\iota_h(\hat{\bm k})$. 
The creation and annihilation operators satisfy
\begin{align}
    \{\hat{c}_{m,\bm k,h},\hat{c}^\dagger_{n,\bm k',h'}\}=\delta_{mn}\delta^3(\bm k-\bm k')\delta_{hh'}, \qquad \{\hat{d}_{m,\bm k,h},\hat{d}^\dagger_{n,\bm k',h'}\}=\delta_{mn}\delta^3(\bm k-\bm k')\delta_{hh'},
\end{align}
which can be consistent with the canonical commutation relation when the following normalization is imposed,
\begin{align}
    |\eta_{n,k,h}(t)|^2+|\lambda_{n,k,h}(t)|^2=1,\label{normalizationcond}
\end{align}
for $\forall t\in\mathbb R$, $\forall k\geq0$ and $h=\pm1$. The Dirac equation reads
\begin{align}
    \ri\frac{d}{dt}\left(\begin{array}{c}\eta_{n,k,h}\\ \lambda_{n,k,h}\end{array}\right)=\left(\begin{array}{cc}
        -\frac{kh}{a} & - M^*_{1/2,n} \\
       - M_{1/2,n} & \frac{kh}{a}
    \end{array}\right)\left(\begin{array}{c}\eta_{n,k,h}\\ \lambda_{n,k,h}\end{array}\right).
\end{align}
Practically, it is useful to decompose the time-dependent mass as
\begin{align}
    M_{1/2,n}=\mu_{n}e^{\ri\Theta_{n}},
\end{align}
with $\mu_n=\left|M_{1/2,n}\right|$, or more explicitly, we can express
\begin{align}
    \cos\Theta_{n}=&\frac{m_{1/2}}{\sqrt{m_{1/2}^2+\left(\frac{n}{Rb}-qg_0\vartheta\right)^2}},\\
    \sin\Theta_{n}=&-\frac{\left(\frac{n}{Rb}-qg_0\vartheta\right)}{\sqrt{m_{1/2}^2+\left(\frac{n}{Rb}-qg_0\vartheta\right)^2}}.
\end{align}
We rewrite the mode equation as
\begin{align}
    \ri\frac{d}{dt}\left(\begin{array}{c}\eta_{n,k,h}\\ \lambda_{n,k,h}\end{array}\right)=\omega_{n,k}\left(\begin{array}{cc}
       \cos\theta_{n,k,h}& \sin\theta_{n,k,h}e^{-\ri\Theta_n} \\
       \sin\theta_{n,k,h}e^{\ri\Theta_n}  & -\cos\theta_{n,k,h}
    \end{array}\right)\left(\begin{array}{c}\eta_{n,k,h}\\ \lambda_{n,k,h}\end{array}\right),
\end{align}
where
\begin{align}
    \omega_{n,k}=\sqrt{\frac{k^2}{a^2}+\mu_n^2}
\end{align}
and $\cos\theta_{n,k,h}=-\frac{kh}{a\omega_{n,k}}$ and $\sin\theta_{n,k,h}=-\frac{\mu_{n,k}}{\omega_{n,k}}$. The mode equation looks like a Schr\"odinger equation of a two-level system. The corresponding instantaneous eigenvectors as
\begin{align}
    {\bm v}^+_{n,k,h}=\left(\begin{array}{c} e^{-\ri\Theta_n}\cos\frac12\theta_{n,k,h}\\ \sin\frac12\theta_{n,k,h} \end{array}\right),\qquad {\bm v}^-_{n,k,h}=\left(\begin{array}{c} -e^{-\ri\Theta_n}\sin\frac12\theta_{n,k,h}\\ \cos\frac12\theta_{n,k,h} \end{array}\right)
\end{align}
with the eigenvalues $\pm\omega_{n,k}$, respectively. We formally write down the solution to the mode equation as
\begin{align}
    \left(\begin{array}{c}\eta_{n,k,h}\\ \lambda_{n,k,h}\end{array}\right)=\gamma_{n,k,h}(t)e^{-\ri\int^t dt'\omega_{n,k}(t')}{\bm v}^+_{n,k,h}+\delta_{n,k,h}(t)e^{+\ri\int^t dt'\omega_{n,k}(t')}{\bm v}^-_{n,k,h},
\end{align}
with auxiliary functions $\gamma_{n,k,h}(t)$ and $\delta_{n,k,h}(t)$ satisfying
\begin{align}
    |\gamma_{n,k,h}(t)|^2+|\delta_{n,k,h}(t)|^2=1,
\end{align}
which follows from the normalization condition~\eqref{normalizationcond}. Substitution of the formal solution to the mode equation reads
\begin{align}
    \dot{\tilde{\gamma}}_{n,k,h}=&-\ri\omega_{n,k}\tilde{\gamma}_{n,k,h}+\frac{\ri}{2}\dot{\Theta}_n\left(1+\cos\theta_{n,k,h}\right)\tilde{\gamma}_{n,k,h}+\frac12\left(\dot\theta_{n,k,h}-\ri\dot\Theta_n\sin\theta_{n,k,h}\right)\tilde{\delta}_{n,k,h},\\
    \dot{\tilde{\delta}}_{n,k,h}=&+\ri\omega_{n,k}\tilde{\delta}_{n,k,h}+\frac{\ri}{2}\dot{\Theta}_n\left(1-\cos\theta_{n,k,h}\right)\tilde{\delta}_{n,k,h}-\frac12\left(\dot\theta_{n,k,h}+\ri\dot\Theta_n\sin\theta_{n,k,h}\right)\tilde{\gamma}_{n,k,h},
\end{align}
where we have defined $\tilde{\gamma}_{n,k,h}=\gamma_{n,k,h}e^{-\ri\int^t dt'\omega_{n,k}(t')}$ and $\tilde{\delta}_{n,k,h}=\delta_{n,k,h}e^{+\ri\int^t dt'\omega_{n,k}(t')}$.

\subsection{Particle production from vacuum}
We briefly review the relation between the auxiliary functions $(\alpha_{n,k}(t),\beta_{n,k}(t))$ and $(\gamma_{n,k,h}(t),\delta_{n,k,h}(t))$ and the particle number density produced from the vacuum. For a scalar field, even if $\beta_{n,k}(t)= 0$ at the initial time, eventually $\beta_{n,k}(t)\neq 0$ due to the nontrivial time-dependence of the background. At a given time $t$, $f_{n,k}(t)$ is generally a linear combination of the positive and the negative frequency mode. Accordingly, the creation and the annihilation operator at $t$ is defined as
\begin{align}
    \hat{A}_{n,\bm k}(t)\equiv& \alpha_{n,k}(t) \hat{a}_{n,\bm k}+\beta^*_{n,k}(t)\hat{b}^\dagger_{n,-\bm k},\\
    \hat{A}^\dagger_{n,\bm k}(t)\equiv& \alpha^*_{n,k}(t) \hat{a}^\dagger_{n,\bm k}+\beta_{n,k}(t)\hat{b}_{n,-\bm k},
\end{align}
which are ``coefficients'' of either positive frequency modes of $\hat{\phi}_n(t,\bm x)$ or negative frequency modes of $\hat{\phi}_n^\dagger(t,\bm x)$. The number density operator of the particle defined at the time $t$ is $\hat{N}_{n,\bm k}(t)=\hat{A}^\dagger_{n,\bm k}(t)\hat{A}_{n,\bm k}(t)$. Evaluating the number density operator in the adiabatic vacuum state $|0\rangle_{\rm in}$ annihilated by $\hat{a}_{n,\bm k},\hat{b}_{n,\bm k}$ for any $\bm k\in {\mathbb R}^3$, we find
\begin{align}
\langle N_{n,\bm k}(t)\rangle  ={}_{\rm in}\langle 0| \hat{A}^\dagger_{n,\bm k}(t)\hat{A}_{n,\bm k}(t) |0\rangle_{\rm in}=\delta^3(\bm k\to 0)|\beta_{n,k}(t)|^2=V_{3D}|\beta_{n,k}(t)|^2,
\end{align}
where $V_{3D}$ denotes the total volume of the 3-space. Therefore, the physical meaning of $|\beta_{n,k}(t)|^2$ turns out to be the phase space number density of the $n$-th KK (anti-)particle having momentum $\bm k$ of $|\bm k|=k$. In an exactly same manner, we find that $|\delta_{n,k}(t)|^2$ can be interpreted as the number density of the spin 1/2 (anti-)particle.

\section{Deformation of the effective potential in extra-natural inflation and KK particle production}
In this section, we show a concrete realization of the extra-natural inflation within our semi-classical 5D QED model. In particular, we focus on the KK particle production that takes place during and after inflation. As we will show below, it is necessary to introduce multiple charged fields to realize an  inflationary potential consistent with the observations, and we find that the KK-Schwinger effect may take place in such a situation.

Let us first show equations of motion and the mode equations. We evaluate the energy-momentum tensor of quantum fields with an adiabatic vacuum state in appendix~\ref{EMtensor}. In our analysis, we take the following assumptions:
\begin{itemize}
    \item The backreaction associated with the particles produced from the vacuum is negligible.
\item We neglect the one-loop corrections originating from  $H_a^2$ and $\dot{H}_a$ in the scalar masses Namely,
\begin{align}
    M_{0,n,A}^2\approx m_{KK,n,A}^2+\frac{m_{0,A}^2}{b}.
\end{align}
\end{itemize}
The first assumption is just for simplicity and will discuss how the dynamics of the inflaton can be modified elsewhere. The second one is a general issue that the scalar one-loop corrections induce higher-curvature corrections, which make the gravitational equations of motion not the second-order but the higher-order. Then, the number of degrees of freedom seems to increase if we solve the equation of motion naively. Neglecting such corrections can be justified when the bulk mass $m_{0,A}^2/b$ is larger than $H_a^2$ and $\dot{H}_a$. Under the above assumptions, we find the time-evolution equations of the classical background fields to be
\begin{align}
   & 3M_{\rm pl}^2H_a^2=\frac{1}{2}\dot\varphi^2+\frac{1}{2}\left(\dot\vartheta+\sqrt{\frac23}\frac{\dot\varphi}{M_{\rm pl}}\vartheta\right)^2+V_{\rm loop}+V_{\Lambda},\label{Hubbleeq}\\
&\ddot{\varphi}+3H_a\dot\varphi-\frac{1}{M_{\rm pl}}\sqrt{\frac32}\left(\dot\vartheta+\sqrt{\frac23}\frac{\dot\varphi}{M_{\rm pl}}\vartheta\right)^2+\partial_\varphi V_{\rm loop}+\partial_\varphi V_{\Lambda}=0,\\
& \ddot\vartheta+3H_a\dot\vartheta-\left(\frac{\sqrt6 H_a\dot\varphi}{M_{\rm pl}}+\sqrt{\frac23}\frac{\ddot\varphi}{M_{\rm pl}}+\frac23\frac{\dot\varphi^2}{M_{\rm pl}^2}\right)\vartheta+\partial_\vartheta V_{\rm loop}=0,
\end{align}
where we have defined the one-loop effective potential
\begin{align}
V_{\Lambda}=&2\pi R \Lambda^5 e^{-\sqrt{\frac23}\varphi/M_{\rm pl}},\\
   V_{\rm loop}= &\sum_{A}\frac{N_A}{2\pi^2L^4b^2}\Biggl[\frac{(LM_{0,A})^{5}}{30}-\sum_{n=1}^{\infty}\cos\left( n q_A g_0\vartheta L \right)e^{-x_{n,0,A}}\frac{3+3x_{n,0,A}+x_{n,0,A}^2}{n^5}\Biggr]\nonumber\\
   &-\sum_B\frac{N_B}{\pi^2L^4b^2}\left[\frac{(Lm_{1/2,B})^{5}}{30}-\sum_{n=1}^{\infty}\cos\left( n q_B g_0\vartheta L\right)e^{-x_{n,1/2,B}}\frac{3+3x_{n,1/2,B}+x_{n,1/2,B}^2}{n^5}\right],\label{Veff}
\end{align}
where $N_{A,B}$ are degeneracies of each field and $x_{n,0,A}\equiv nLM_{0,A}$, $x_{n,1/2,B}\equiv nLm_{1/2,B}$. We note that the overall scale of the potential can be tuned by the following parameter rescaling: With a constant $C$, we make $R\to C^{-1} R$, $m_{0,A}\to Cm_{0,A}$, $m_{1/2,B}\to Cm_{1/2,B}$, $g_0\to Cg_0$, $2\pi R\Lambda^5\to  2\pi R\Lambda^5C^4$, which change the overall scale of the effective potential $V\to C^4V$. We also note that the first term of each square bracket originates from the zero mode and is the (finite) correction to the bulk cosmological constant $\Lambda^5$.

Let us consider the dynamics of matter fields. The scalar mode functions can be rewritten in terms of three real quantities $N^{(0)}_{n,k}\equiv |\beta_{n,k}|^2=|\tilde{\beta}_{n,k}|^2$, $R^{(0)}_{n,k}\equiv {\rm Re}(\tilde{\alpha}_{n,k}\overline{\tilde\beta}_{n,k})$ and $I^{(0)}_{n,k}\equiv{\rm Im}(\tilde{\alpha}_{n,k}\overline{\tilde\beta}_{n,k})$, which satisfy
\begin{align}
    \frac{d}{dt}N^{(0)}_{n,k,A}=&\frac{\dot\omega_{0,n,k,A}}{\omega_{0,n,k,A}}R^{(0)}_{n,k,A},\label{dN0}\\
    \frac{d}{dt}R^{(0)}_{n,k,A}=&\frac{\dot\omega_{0,n,k,A}}{2\omega_{0,n,k,A}}\left(1+2N^{(0)}_{n,k,A}\right)-2\omega_{0,n,k,A}I^{(0)}_{n,k,A},\label{dR0}\\
    \frac{d}{dt}I^{(0)}_{n,k,A}=&2\omega_{0,n,k,A}R^{(0)}_{n,k,A}\label{dI0}
\end{align}
that follow from the mode equations~\eqref{tildeab}.
For a spinor field, we define $N^{(1/2)}_{n,p,h}=|\delta_{n,p,h}|^2$, $R^{(1/2)}_{n,p,h}={\rm Re}(\tilde\gamma_{n,p,h}\overline{\tilde{\delta}}_{n,p,h})$ and $I^{(1/2)}_{n,p,h}={\rm Im}(\tilde\gamma_{n,p,h}\overline{\tilde{\delta}}_{n,p,h})$, which follow
\begin{align}
    \frac{d}{dt}N^{(1/2)}_{n,k,h,B}=&-\dot\theta_{n,k,h,B}R^{(1/2)}_{n,k,h,B}+\dot\Theta_{n,B}\sin\theta_{n,k,h,B}I^{(1/2)}_{n,k,h,B},\label{dN12}\\
    \frac{d}{dt}R^{(1/2)}_{n,k,h,B}=&(2\omega_{n,k,B}-\dot\Theta_{n,B}\cos\theta_{n,k,h,B})I_{n,k,h,B}+\dot\theta_{n,k,h,B}\left(2N^{(1/2)}_{n,k,h,B}-1\right),\label{dR12}\\
    \frac{d}{dt}I^{(1/2)}_{n,k,h,B}=&-(2\omega_{n,k,B}-\dot\Theta_{n,B}\cos\theta_{n,k,h,B})R^{(1/2)}_{n,k,h,B}-\dot\Theta_{n,B}\sin\theta_{n,k,h,B}\left(2N^{(1/2)}_{n,k,h,B}-1\right).\label{dI12}
\end{align}
In general, it is difficult to solve the above equations analytically, and one needs numerical simulations. Nevertheless, we will give an analytic estimate for the behavior of the particle number density. We note that, using the full expressions of energy-momentum tensors and the current, one can also numerically simulate the dynamics including the backreaction effects, which we will neglect in this work and show that such effects are negligible for our choice of parameters shown later. However, in general, the backreaction effects can become important for inflationary dynamics and for (p)reheating. 

\subsection{Tuning potential and multiple charged fields}\label{tuning}
The one-loop effective potential~\eqref{Veff} generally depends on both the radion field $\varphi$ and the Wilson line field $\vartheta$, but the contributions from the cosmological constant and the one-loop effective potential due to the neutral fields depend only on $\varphi$. Here, for simplicity, we consider the case that the radion is stabilized by the neutral sector and treat the value of $b$  (accordingly $L$) as a constant parameter. Furthermore, although the $\vartheta$-dependent term appears as an infinite series, the large $n$ contributions are suppressed either by $1/n^5$ or the exponential factor for massive cases. Therefore, we may approximately treat it by the leading two terms $n=1$, which is a good approximation. 

Within the extra-natural inflation, the effective decay constant of $\vartheta$
\begin{align}
    f_{\rm eff}=\frac{1}{g_0L}
\end{align}
can be larger than 4D Planck scale when $g_0\ll1$ as discussed in \cite{Arkani-Hamed:2003wrq,Arkani-Hamed:2003xts} even when the extra dimension volume $L$ is large enough. Then, a large field inflation can be realized, which reduces the fine-tuning of the initial condition. If there is only a single charged sector, the potential is effectively determined by a single sinusoidal function since the higher-order contributions are suppressed and $n=1$ contribution dominates.\footnote{When the bulk mass is zero, the exponential factor is negligible but simply suppressed by the $n^5$ factor in the denominator. For the modes with bulk masses, the exponential suppression and the $n^5$ factor suppress the contributions from $n>1$.} However, the simplest one-parameter natural inflation model cannot be compatible with the Planck 2018 result~\cite{Planck:2018jri} and the latest result from Atacama Cosmology Telescope Data Release 6~\cite{ACT:2025fju,ACT:2025tim} combined with Planck, BICEP/Keck and Baryon Acoustic Oscillation (BAO) data. Therefore, despite a ``natural'' realization of an effective large decay constant, the simplest case cannot explain the observational data.\footnote{Note that we should take into account the BAO-CMB tension reported in \cite{Ferreira:2025lrd}. The shift of the preferred value of the scalar perturbation spectral index $n_s$ seems to originate from the discrepancies of the central values of the matter energy $\Omega_m$ and the sound horizon times the Hubble parameter today $r_d h$ by the CMB observation (Planck, ACT, SPT) and the BAO measurement (DESI). The recent data analysis of South Pole Telescope (SPT) also exhibits a similar feature~\cite{SPT-3G:2025bzu}. From a conservative viewpoint, it remains unclear if the spectral index of the combined result of CMB and BAO should be respected. Nevertheless, in either case, the single sinusoidal potential is not preferred anyway, and the deformation of effective potential would be necessary independently of the BAO-CMB tension.} Therefore, it is necessary to deform the simplest effective potential.

What kind of modification is necessary? There is a possibility to deform the potential of the Wilson line field $\vartheta$: We may introduce multiple U(1) charged fields having different masses and charges, which realize the so-called multi-natural inflation model~\cite{Czerny:2014wza,Czerny:2014xja,Higaki:2014pja,Higaki:2014mwa}. Roughly speaking, addition of different frequency modes may realize an inflationary plateau, although each contribution itself does not. As we will show in the illustrating model, it is possible to tune the potential by introducing charged particles with different charges. We do not argue that an inflationary potential compatible with cosmological data can be naturally realized within our framework. What we argue here is the possibility to make the inflationary model compatible with the latest data.

To illustrate the situation, let us allow for any fine-tunings to the mass parameters as well as the degeneracy of each particle. Since the radion stabilization can be achieved by the neutral sector with a cosmological constant, we treat the value of $b$ as a parameter. Then, the effective potential can be written as
\begin{align}
    V_{\rm loop}\approx C-\sum_{A}C_A\cos\left(\frac{q_A\vartheta}{f_{\rm eff}}\right)+\sum_{B}C_B\cos\left(\frac{q_B\vartheta}{f_{\rm eff}}\right)
\end{align}
where
\begin{align}
    C_A=&\frac{N_A(3+3x_{1,0,A}+x_{1,0,A}^2)e^{-x_{1,0,A}}}{2\pi^2L^4b^2},\\
C_B=&\frac{N_B(3+3x_{1,1/2,B}+x_{1,1/2,B}^2)e^{-x_{1,1/2,B}}}{\pi^2L^4b^2}.
\end{align}
If there are various charged particles with various bulk masses, the effective potential is given by linear combinations of various sinusoidal functions, which is the case of the ``multi-natural inflation''~\cite{Czerny:2014wza,Czerny:2014xja,Higaki:2014mwa,Higaki:2014pja}. Even for the case with two cosine potential terms, various patterns of potential are realized~\cite{Czerny:2014wza}, which lead to various values of the spectral index and the tensor-to-scalar ratio.

Let us discuss the relation between the multi-charge extension for the deformation of the potential and the KK-Schwinger effect. First of all, the KK-Schwinger effect takes place when $M_{KK,n}^2=0$ where the $n$-th mode temporarily becomes a zero mode. If there is only a single charged sector having a bulk mass smaller than the KK scale $\sim L^{-1}$, no KK Schwinger effect is expected: When the potential is determined by a single sinusoidal function, the field variation associated with inflation is at most over the half period $\vartheta\in [-\frac{\pi}{2q_0g_0 L},\frac{\pi}{2q_0g_0 L}]=[-\frac{1}{4q_0g_0 Rb},\frac{1}{4q_0g_0 Rb}]$. Then, the KK mass of the field having the charge $q_0$ varies as
\begin{align}
    \frac{(|n|-1/4)^2}{R^2b^3}<m^2_{KK,n,q_0}<\frac{(|n|+1/4)^2}{R^2b^3}
\end{align}
for any $n\neq 0$. Note that, for $n=0$ mode, the ``KK mass'' varies over $0\leq m^2_{KK,0,q_0}<\frac{1}{16R^2b^3}$.  Then, only $n=0$ mode experiences $m^2_{KK,n,q_0}=0$, and therefore, the production rate of $|n|>1$ modes would be exponentially suppressed by KK masses. In such a case, only the preheating production of the zero mode can occur.

The presence of multiple charged particles changes the situation and generally we expect the KK Schwinger effect: Suppose the potential is dominated by the sinusoidal function by $q_0$-sector as above (but with deformation by contributions from other charged fields). If there is a charged particle with the charge $q_1$ satisfying $|q_1|>4|q_0|$. In such a case, the KK mass of such a field changes as
\begin{align}
    \frac{(|n|-|q_1/(4q_0)|)^2}{R^2b^3}<m^2_{KK,n,q_1}<\frac{(|n|+|q_1/(4q_0)|)^2}{R^2b^3}.
\end{align}
By assumption, at least either $n= 1$ or $n=-1$ mode experiences vanishing of the KK mass $m^2_{KK,\pm1,q_1}=0$ at some time. At the time of vanishing of the KK mass, $n=\pm1$ modes can be created without the exponential suppression by the ``heavy KK mass''. If the charge ratio is large enough, $|n|>1$ modes can also be produced during and after inflation. Thus, in the presence of various charges, the KK-Schwinger effect generally occurs. We note that the KK modes of fields having bulk masses larger than the KK scale $L^{-1}$ give no contributions to the potential, and the production rate of such particles is exponentially suppressed. Then, we may simply neglect such fields except for the zero mode contribution to the effective potential.\footnote{Note, however, that the one-loop effect of such heavy zero modes just change the bulk cosmological constant, which can be renormolized by the bare parameter, and therefore, the fields having very large bulk mass are essentially decoupled.}

\subsection{Illustrative model}\label{illustrative}
We show an illustrative toy model that achieves both inflation and radion stabilization. We choose the set of parameters as
\begin{align}
    R=&40\times C^{-1}, \ g_0=0.02C,\nonumber\\
    2\pi R\Lambda^5=&1.1182\times 10^{-3}C^4,\nonumber
\end{align}
\begin{align}
    (N_A,m_{0,A},q_A)=&(28,1.8\times10^{-1}C,0), \quad (2,5\times10^{-4}C,1)\quad (2,2.014\times10^{-4}C,5)\nonumber\\
    (N_B,m_{1/2,B},q_B)=&(16,1.9\times 10^{-1}C,0),\quad (1,5.2\times10^{-4}C,1),\quad (1,2\times10^{-4}C,5).\nonumber
\end{align}
Here and hereafter, the unit of mass is the 4D Planck mass $M_{\rm pl}=1$ where $M_{\rm pl}\sim 2.4\times10^{18}$~GeV. A constant factor $C=0.378$ is introduced to tune the height of the potential (see the comments below \eqref{Veff}). Here, we have introduced the particles neutral under the U(1) gauge symmetry, which only contribute to the radion stabilization, and an effectively supersymmetry-like spectrum for the charged sector such that the hierarchy between the radion mass and the inflaton potential is realized. We should emphasize that this model is nothing more than just an illustration of KK particle production, and we have allowed ourselves the fine-tuning of model parameters. 

\begin{figure}
    \centering
    \includegraphics[width=0.7\linewidth]{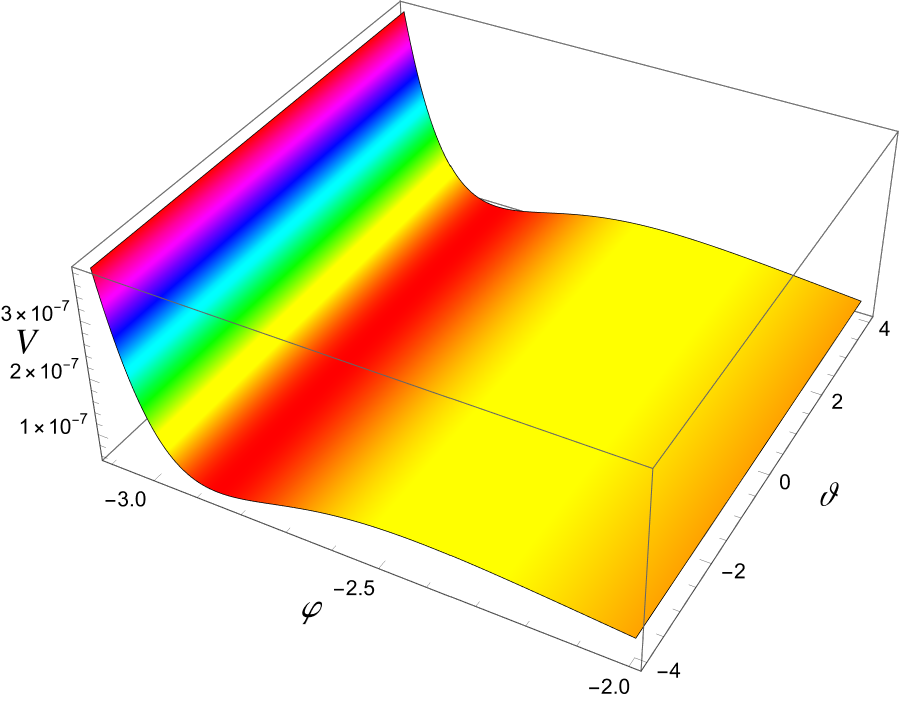}
    \caption{The effective potential of the radion and the Wilson line modulus $(\varphi,\vartheta)$. The potential looks flat along $\vartheta$ due to the hierarchy of the effective mass of $(\varphi,\vartheta)$. During inflation driven by $\vartheta$, the radion $\varphi$ can be effectively treated as a constant.}
    \label{fig:3D potential}
\end{figure}

\begin{figure}
    \centering
    \includegraphics[width=0.8\linewidth]{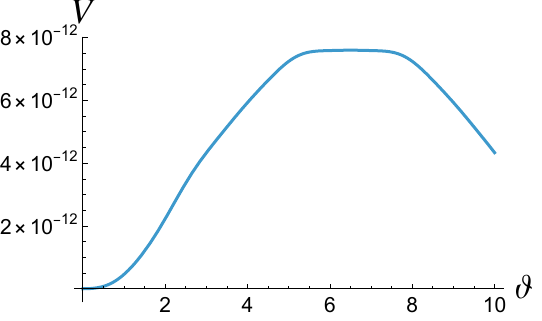}
    \caption{The potential of $\vartheta$ with a fixed $\varphi$ at the bottom.}
    \label{fig:1D potential}
\end{figure}

Let us first show that the radion stabilization is achieved without affecting the inflation dynamics. In Fig.~\ref{fig:3D potential}, the hierarchy of the potential can be seen. One can confirm that the radion does not contribute much to the inflationary dynamics. Hereafter, we will fix the radion and consider only the dynamics of $\vartheta$. At the minimum, the physical length of the compact space radius is
\begin{align}
    \langle L\rangle =2\pi R \langle b\rangle=63.48
\end{align}
and the size of the compact space is larger than the 4D Planck length. 
Let us look at the 1D effective potential of $\vartheta$ shown in Fig.~\ref{fig:1D potential}.\footnote{ In deriving the effective potential, we have used the following approximation: The potential minimum of the one given in Fig.~\ref{fig:3D potential} is indeed positive and its size is smaller than the typical scale of $V$ but yet larger than the observed value $10^{-120}$. Therefore, we have subtracted the potential as $V-V_{\rm min}$ where $V_{\rm min}$ is the minimum value. We could have fine-tuned the 5D cosmological constant $\Lambda^5$ such that $V_{\rm min}$ but it requires an unnecessary effort for searching the parameter $\Lambda^5$, which is possible in principle, but would give no physical significance.} Now, the potential minimum is at $\vartheta=0$, as we have assumed. 
The potential shown in Fig.~\ref{fig:1D potential} is no longer just a sinusoidal form but has a plateau region, which never appears if the one-loop contribution from a single charged sector dominates the effective potential. We should emphasize again that the plateau is just accidental and not a general feature of the one-loop effective potential from multi charged particles.

\begin{figure}[tbp]
  \begin{minipage}[b]{0.48\columnwidth}
    \centering
    \includegraphics[width=\columnwidth]{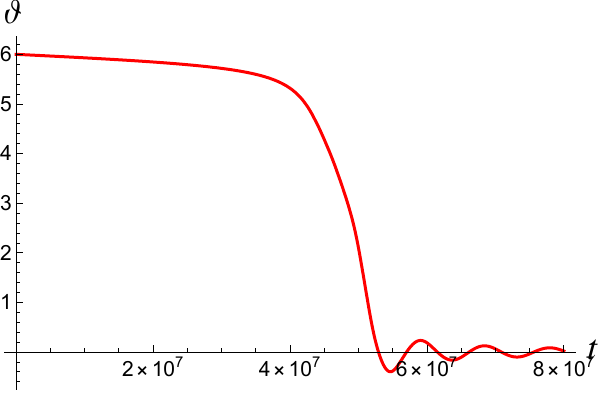}
    \caption{The time evolution of $\vartheta$.}
    \label{fig:varphi}
  \end{minipage}
  \hspace{0.04\columnwidth} 
  \begin{minipage}[b]{0.48\columnwidth}
    \centering
    \includegraphics[width=\columnwidth]{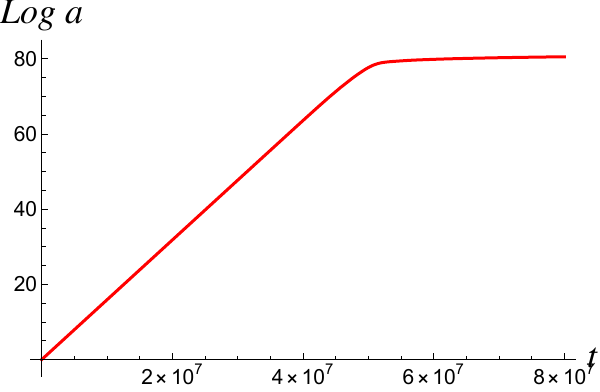}
    \caption{The time evolution of the scale factor.}
     \label{fig:N}
  \end{minipage}
\end{figure}

\begin{figure}[tbp]
  \begin{minipage}[b]{0.48\columnwidth}
    \centering
    \includegraphics[width=\columnwidth]{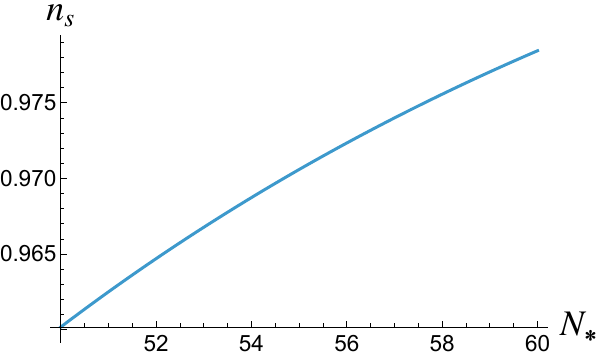}
    \caption{The scalar spectral index $n_s$ as the function of the corresponding e-foloding number $N_*$ of the horizon exit.}
    \label{fig:ns}
  \end{minipage}
  \hspace{0.04\columnwidth} 
  \begin{minipage}[b]{0.48\columnwidth}
    \centering
    \includegraphics[width=\columnwidth]{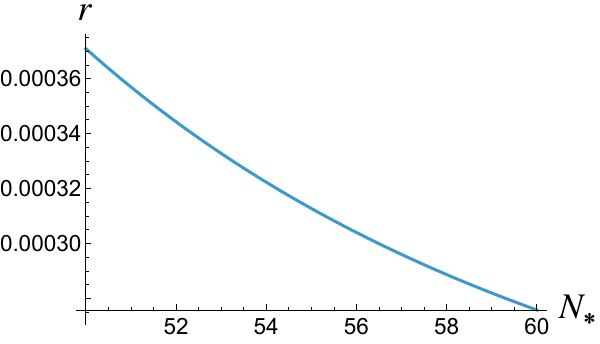}
    \caption{The tensor-to-scalar ratio $r$ as the function of the corresponding e-folding number $N_*$ of the horizon exit.}
     \label{fig:r}
  \end{minipage}
\end{figure}

We have numerically solved the dynamical equations, and the results are as follows: In Figs.~\ref{fig:varphi},\ref{fig:N}, we have shown the time evolution of $\varphi$ and $\log a(t)=N$, where we have taken the initial conditions $\vartheta(0)=6$ and $\dot\vartheta(0)=0$. The spectral index of the scalar perturbation $n_s$ and the tensor-to-scalar ratio $r$ are expressed as $n_s=1+2\eta-6\epsilon$, $r=16\epsilon$, where the slow-roll parameters $\epsilon,\eta$ are
\begin{align}
    \epsilon=\frac{1}{2}\left(\frac{\partial_\vartheta V}{V}\right)^2,\qquad \eta=\frac{\partial_\vartheta^2V}{V}.
\end{align}
We show the values of $n_s$ and $r$ as the function of the corresponding e-folding number $N_*=\log a_*$ at the horizon exit in Figs.~\ref{fig:ns}, \ref{fig:r}. We find that the value of $r$ is consistent with the current bound. Also, the value of $n_s$ can be consistent with both the Planck and the ACT depending on the choice of $N_*$, which is determined by the reheating after inflation.\footnote{We have fixed the normalization of the potential height such that $P_\zeta\sim 2\times10^{-9}$ for $50\leq N_*\leq 60$. One has to completely fix the potential height such that the correct amplitude of the power spectrum is reproduced, which slightly changes the values of particle masses, the volume of the compact direction and so on.  } 

We should emphasize that the consistency with the current observational data is not essential. What we argue here is that the multi-natural extension of the extra-natural inflation is a necessary (but not sufficient) condition to realize an effective inflationary potential compatible with the present cosmological data. More generally, we may introduce more charged particles, which leads to a richer structure and may reduce the fine-tuning of parameters, as we will briefly discuss in Sec.~\ref{conclusion}.

\subsection{KK particle production via KK-Schwinger effect}\label{KKSchwinger}
\begin{figure}
    \centering
    \includegraphics[width=0.8\linewidth]{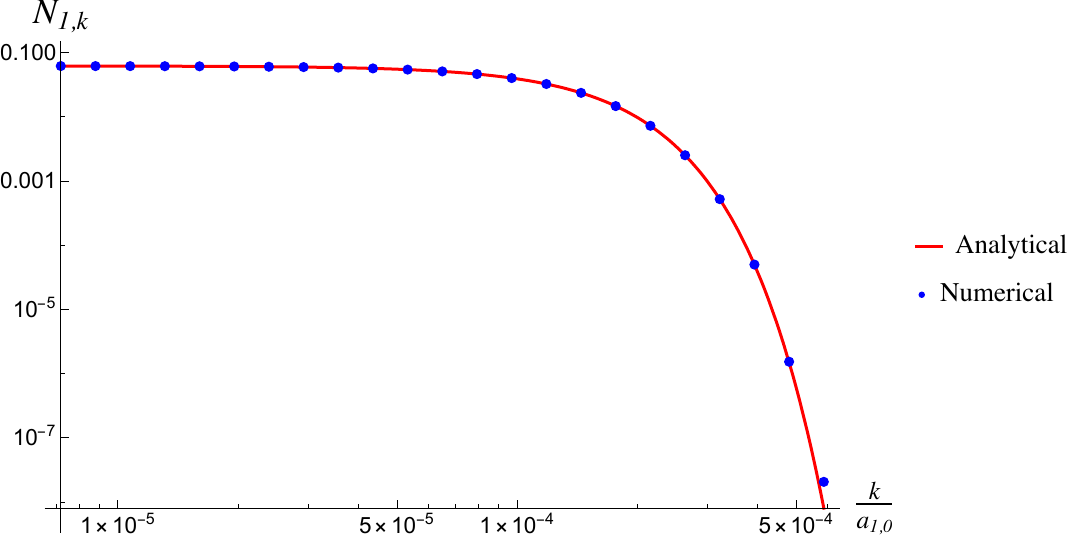}
    \caption{The phase space number density $N_{n,k,A}(t_f)$ for the $n=1,q=5$ mode of a scalar field. The red curve is the analytic formula \eqref{analyticKK} with numerically derived parameters $t_{n,A}\approx $ and $\dot\vartheta|_{t=t_{n,A}}\approx $. The blue dots are numerical results from the time-evolution equations~\eqref{dN0}-\eqref{dI0}. The correspondence between the analytic and the numerical results are excellent. Note that the horizontal axis is the momentum with the scale factor at the time of the $n=1$ mode production $t\sim 3.7\times 10^7$.}
    \label{fig:N1density}
\end{figure}

\begin{figure}
    \centering
    \includegraphics[width=0.8\linewidth]{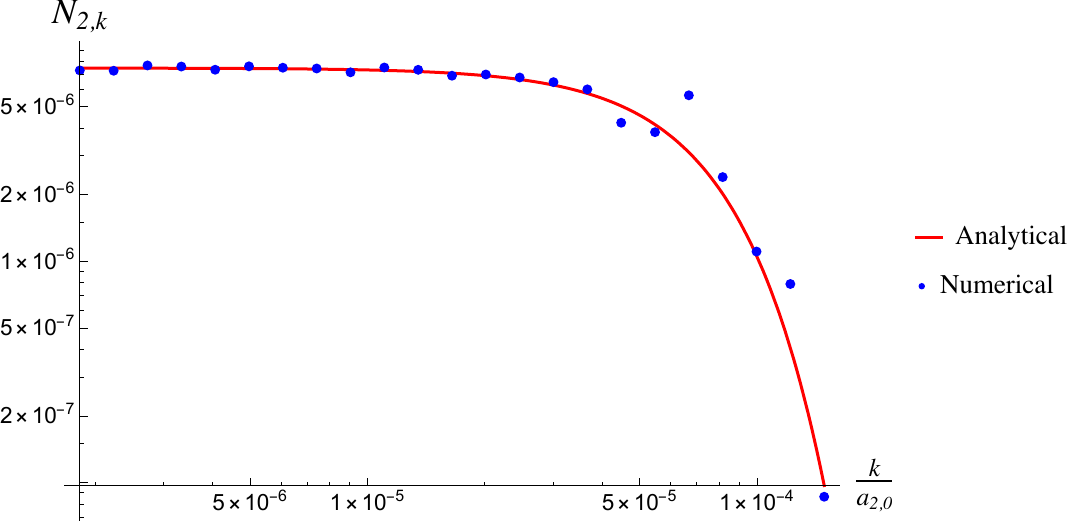}
    \caption{The phase space number density $N_{n,k,A}(t_f)$ for the $n=2,q=5$ mode of a scalar field. The red curve is the analytic formula \eqref{analyticKK} with numerically derived parameters $t_{n,A}\approx $ and $\dot\vartheta|_{t=t_{n,A}}\approx $. The blue dots are numerical results from the time-evolution equations~\eqref{dN0}-\eqref{dI0}. Compared with the $n=1$ case, we find small discrepancy between the numerical result and the analytic one, which seems due to the accuracy of the numerical calculations.}
    \label{fig:N2density}
\end{figure}

\begin{figure}
    \centering
    \includegraphics[width=0.8\linewidth]{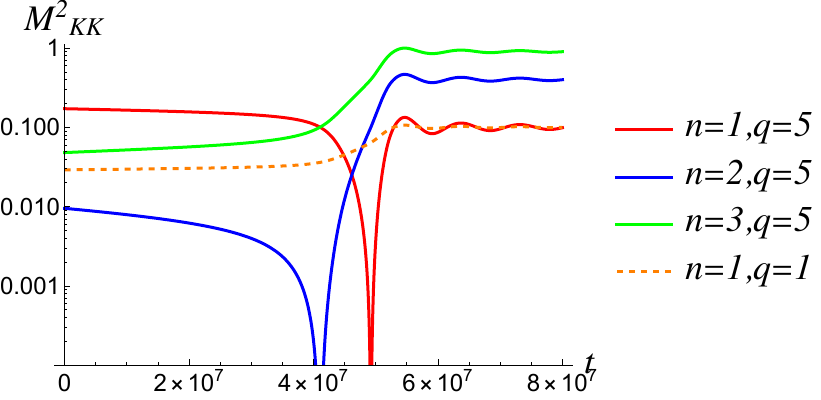}
    \caption{Time dependence of $M^2_{KK,n}$ for $n=1,2,3$, $q=1,5$. For $q=5,n=1,2$ modes, the KK mass vanishes at some time whereas others are non-vanishing. The KK-Schwinger effect occurs only for the modes with vanishing KK masses.}
    \label{fig:KKmass}
\end{figure}

Let us show that KK particles are indeed produced from the vacuum. In the following discussion, we focus only on scalar fields, but the behavior of spinor fields is almost the same. We have numerically solved the set of equations~\eqref{dN0}-\eqref{dI0} for scalar KK modes having $q=5,n=1,2$. As explained in Sec.~\ref{tuning}, charged particles having $q=1$ cannot be produced for any KK number $n$. In Figs.~\ref{fig:N1density}, \ref{fig:N2density}, we have shown the final particle number density as a function of $k$ for $n=1,2$. (The analytic formula is explained below.) Although we have not shown here, one can confirm that the production rates of $q=1,n\geq1$ modes are negligibly small for any $k$. Let us first clarify why only $n=1,2$ and $q=5$ are produced whereas others are not. In Fig.~\ref{fig:KKmass}, we have shown the time-dependence of the KK mass, which shows that KK masses of $n=1,2$,$q=5$ vanish at some time, but others ($n=3,q=5$ or $n=1,q=1$) do not. Figure~\ref{fig:KKmass} also clarifies why the KK particles can be efficiently produced: As we explained, the KK particles become a ``zero mode'' at the time of their production and get heavy again as $\vartheta\to 0$. Namely, the KK particles can be produced however heavy they will become.\footnote{We have implicitly assumed that the bulk mass is small enough. More precisely speaking, if a charged particle has a very large bulk mass, they would not be produced from the vacuum. Also, their contributions to the one-loop effective potential becomes zero (except the $\vartheta$-independent terms), which means that such heavy particles decouple from the theory.} As shown in Fig.~\ref{fig:KKmass}, the KK mass scale eventually becomes $M_{KK,n}\to  0.1\times n^2\sim 2.4\times 10^{17} n^2 $ GeV, which is much higher than the Hubble scale during inflation $H_{\rm inf}\sim 1.8\times 10^{-6}\sim 4.3\times 10^{12} $ GeV. This hierarchy between the Hubble scale and the produced KK mass is one of the differences of the KK-Schwinger effect from the cosmological particle production of superheavy particles~\cite{Kuzmin:1998uv,Chung:1998zb,Chung:1998ua,Kolb:1998ki,Chung:2001cb,Ling:2025nlw}, where the particle mass should be less than or equal to $H_{\rm inf}$ for the efficient production. One may also think of the Hawking radiation of the superheavy particles due to the ``thermal bath'' of de Sitter spacetime, where the temperature is $T_{\rm dS}\sim 2\pi H_{\rm inf}$. The Boltzmann factor $e^{-m_{\rm KK}/(2\pi H_{\rm inf})}$ is too small, and therefore, the gravitational particle production mechanism cannot produce the KK modes, which shows the difference between the cosmological particle production mechanism and the KK-Schwinger effect.

We should emphasize that the backreaction effects are neglected in the above arguments, and the backreactions to $\vartheta$ prevent the KK mass from getting larger in general. Physically speaking, the KK pair production induces a ``current'' along the compact space, which shields the electric field. In terms of the 4D effective theory, the energy density of the produced KK particles depends on $\vartheta$ and behaves as an effective potential for $\vartheta$. As the KK mass of the produced mode becomes large, the energy density of the KK mode increases, which causes the force on $\vartheta$ to decrease the mass energy.\footnote{Such backreaction forces yield interesting phenomena called moduli trapping~\cite{Kofman:2004yc,Kikuchi:2023uqo}.} In our example, as we will estimate below, the energy density of the produced particles is much smaller than the energy density of $\vartheta$, and therefore, we can safely neglect the backreaction.

We analytically estimate the KK particle number density produced via the KK-Schwinger effect during and after inflation. A systematic way to quantify the particle production associated with nontrivial backgrounds can be achieved by using the (exact) WKB analysis and analyzing the Stokes phenomena~\cite{Dumlu:2010ua,Dabrowski:2014ica,Dabrowski:2016tsx,Li:2019ves,Enomoto:2020xlf,Taya:2020dco,Hashiba:2021npn,Sou:2021juh,Hashiba:2022bzi}. However, since we are interested in the case where a KK mode experiences the production event only once, we do not need such a sophisticated method but just use a well-known result: The one-particle energy of a complex scalar reads
\begin{align}
    \omega_{0,n,k,A}^2\approx \frac{k^2}{a^2}+\frac{m_0^2}{b}+\frac{1}{b^3}\left(\frac{n}{R}-q_Ag_0b\vartheta\right)^2,
\end{align}
where we have dropped the Hubble-induced terms as we have assumed. Note that only the inflaton field $\vartheta$ is time-dependent since the scale factor changes more slowly. We define $t_{A,n}$ as the time when $\frac{n}{R}-q_Ag_0b\vartheta(t_{A,n})=0$. Around $t=t_{A,n}$ we may approximate the one-particle energy as
\begin{align}
    \omega_{0,n,k,A}^2(t)\approx (\omega_{0,n,k,A}^{\rm min})^2+v_{n,A}^2(t-t_{n,A})^2
\end{align}
where we have defined
\begin{align}
\omega_{0,n,k,A}^{\rm min}\equiv &\sqrt{\frac{k^2}{a^2(t_{n,A})}+\frac{m_0^2}{b}},\\
   v_{n,A}\equiv &q_Ag_0b^{-\frac12}\dot\vartheta|_{t=t_{n,A}}.
\end{align}
Then, around $t=t_{A,n}$, the mode equation can be approximated by the parabolic cylinder function, and we can apply the known formula of the produced particle-number density (see e.g.~\cite{Kofman:1997yn}),
\begin{align}
    N_{n,k,A}(t)\approx \left\{\begin{array}{c}0\qquad (\text{ for } t<t_{n,A})\\
    \exp\left(-\frac{\pi (\omega_{0,n,k,A}^{\rm min})^2}{v_{n,A}}\right)\qquad (\text{for }t>t_{n,A})\ .\end{array}\right. \label{analyticKK}
\end{align}
Note that the production rate of spinor fields by a single particle production event is almost the same as discussed in \cite{Peloso:2000hy}.If there were multiple KK particle production events, Bose enhancements and Pauli blocking distinguish the behavior of the bosonic and fermionic particle number density, but the resonant effects do not occur unless we assume a particle having a very large U(1) charge. So, the production rate formula above can be applied to both scalars and spinors in the same way. As we explicitly show in Figs.~\ref{fig:N1density},\ref{fig:N2density} the correspondence between the above analytic formula and the numerical results is very good. In other words, we can reliably use the above analytic formulas in analytically estimating the physical quantities such as the energy density of the produced KK particles. Nevertheless, for the $n=2$ case, we find a small discrepancy between numerical and analytical results. We believe that the small discrepancy originates from numerical errors.

We should emphasize that our estimation given here works only when $M_{KK,n,A}^2=0$ once. If $M_{KK,n,A}^2=0$ is satisfied twice, the secondary (or later) particle production event has ``interference effects'' and the estimation of the particle number density is more involved, which is the same as the stochastic resonance regime of preheating~\cite{Kofman:1997yn}. Note also that the zero mode would experience the standard preheating by $\vartheta$ oscillating around $\vartheta=0$.\footnote{We have assumed that the inflaton $\vartheta$ falls down to $\vartheta=0$, and therefore defined $n=0$ as a zero mode. If $\vartheta\neq 0$ at a minimum, we should relabel the KK number $n$ and identify the lowest mass mode as a zero mode.}

\subsection{Fate of super heavy KK particles}\label{fateofKK}
We briefly discuss the fate of KK particles produced via the KK-Schwinger effect. In particular, due to the KK number conservation, the lowest KK modes $n=\pm1$ cannot decay into zero modes, which makes the produced KK particle a candidate for superheavy dark matter. It is also possible that the KK particles overcloses the universe, which is a new problem within the higher-dimensional theory.

The analytic expression of the particle number density enables us to derive the analytic formula of the total energy density of the produced KK particle: Dropping oscillatory factors $R_{n,k,A}^{(0)},I_{n,k,A}^{(0)}$,\footnote{Let us explain why $R_{n,k,A}^{(0)},I_{n,k,A}^{(0)}$ become oscillatory factors. In the limit $\dot\omega_{0,n,k,A}\to 0$, $N_{n,k,A}^{(0)}$ becomes a constant following from \eqref{dN0}. Also, \eqref{dR0},\eqref{dI0} read $\ddot{R}_{n,k,A}^{(0)}\sim -4\omega_{0,n,k,A}^2R_{n,k,A}^{(0)}$ where we have dropped terms with $\dot\omega_{0,n,k,A}$. Then, the solutions are $R_{n,k,A}^{(0)},I_{n,k,A}^{(0)}\sim e^{\pm 2\ri \omega_{0,n,k,A}t}$. Thus, the asymptotic behavior of them are just oscillatory factors, which would become zero on average. } the energy density of $\phi_{n,A}$ can be estimated as 
\begin{align}
    \rho_{n,A}\approx& \int \frac{d^3\bm k}{(2\pi)^3a^3(t)}2\omega_{0,n,k,A}N_{n,k,A}^{(0)}(t)
    \nonumber\\
    \approx& \int \frac{d^3\bm k}{(2\pi)^3a^3(t)}2\omega_{0,n,k,A}\exp\left[-\frac{\pi (k^2/a(t_{n,A})^2+m_0^2/b)}{v_{A,n}}\right]\theta_H(t-t_{n,A})\nonumber\\
    =&\frac{v_{n,A}}{4\pi^3}\left(\frac{a_{n,A}}{a} \right)^2e^{-\frac{m_{0,A}^2}{bv_{n,A}}}e^{-\frac{\pi M_{0,n,A}^2}{2v_{n,A}}\frac{a^2}{a^2_{n,A}}} M_{0,n,A}^2K_1\left(\frac{\pi M_{0,n,A}^2}{2v_{n,A}}\frac{a^2}{a^2_{n,A}}\right)\theta_H(t-t_{n,A})\nonumber\\
    =& \frac{1}{4\pi^3}\left(\frac{a_{n,A}}{a}\right)^3v_{A,n}^\frac32 M_{KK,n,A}\exp\left(-\frac{\pi m_{0,A}^2}{b v_{A,n}}\right)\left(1+\cdots\right),\label{energy density formula}
\end{align}
where $a_{n,A}=a(t_{n,A})$, and $\theta_H(t)$ denotes the Heaviside step function.
The first approximate equality means that we have dropped the oscillatory contributions as well as the Hubble-induced corrections. The second one means the approximate time-dependence of $N_{n,k,A}^{(0)}$~\eqref{analyticKK}. The last expression is the one leading order in $M_{KK,n,A}$ and the ellipses denote $\mathcal{O}\left(\frac{m_{0,A}^2}{M_{KK,n,A}^2}\right)$ or $\mathcal{O}\left(\frac{v_{n,A}}{M_{KK,n,A}^2}\right)$. 
\begin{figure}
    \centering
    \includegraphics[width=0.8\linewidth]{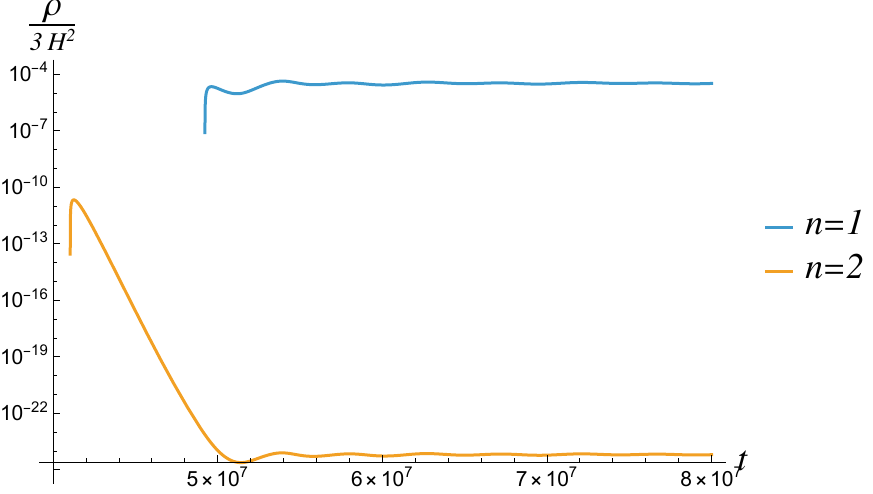}
    \caption{The ratio of the total energy density and the energy density of $n=1,2$ modes based on the formula~\eqref{energy density formula}. We have substituted numerically derived values of $v_{n,A},\ t_{n,A},$ and $a_{n,A}$. Since the $n=2$ mode is produced during inflation, its energy density decays exponentially in time, whereas the $n=1$ mode is produced after inflation. This shows that the $n=2$ mode can be negligible. }
    \label{fig:energy ratio1}
\end{figure}
\begin{figure}
    \centering
    \includegraphics[width=0.8\linewidth]{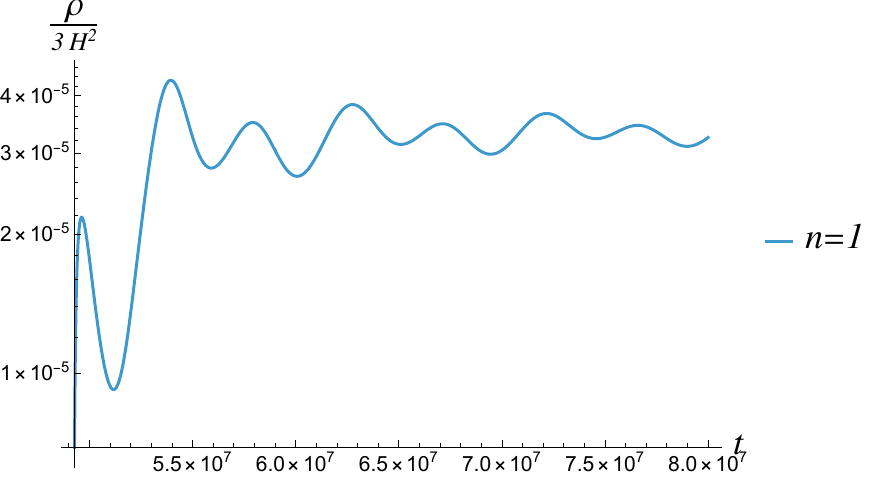}
    \caption{The ratio of the total energy density and the energy density of $n=1$ mode based on the formula~\eqref{energy density formula}. This shows that the KK particle energy density is $\mathcal{O}(0.001)\%$ and its backreaction effect would be consistently negligible.}
    \label{fig:energy ratio2}
\end{figure}

We have shown the ratio between the produced particle energy density and the total energy density $3H^2$ in Figs.~\ref{fig:energy ratio1}, \ref{fig:energy ratio2}. One finds that the $n=2$ mode is produced during inflation (or more precisely near the end of inflation), and therefore, the energy density exponentially decays in time. On the other hand, the $n=1$ mode is produced after inflation, which is why it remains as $\mathcal{O}(0.001)\%$ of the energy density. 


Within our setup, the lowest KK particles $n=\pm1$ are quite stable due to the KK number conservation originating from the translational invariance of the compact space. The relic abundance of such stable particles should not exceed the observed value of the dark matter energy density today. In particular, due to superheavy mass of the KK particles, their relic density has to be extremely small. For precise estimation of the allowed parameter regions, we need to specify how the standard model particles are coupled to the KK particles and the inflaton $\vartheta$. We will give a few comments about the modification of our model to include the standard model sector at the end of this subsection. Here, we tentatively assume that the inflaton $\vartheta$ couples to the standard model particles and completes reheating at the temperature $T_R$ and discuss the constraints on the allowed KK particle energy density. We assume that the KK particles are completely stable, and the inflaton decays only to the standard model sector, which is reasonable within our setup since the inflaton is lighter than the zero modes of the charged particles. Until the inflaton decay, the KK particle and the inflaton energy density ratio is approximately constant as both are matter-like. We define the energy density ratio as
\begin{align}
    r_{KK}\equiv \frac{\rho_{KK}}{\rho_{\rm inf}}\Biggl|_{\rm R}=\frac{\rho_{KK}}{3H^2}\Biggl|_{\rm R}
\end{align}
where $\cdot \ |_R$ means that it is evaluated before reheating but well after the time that the KK mass becomes constant.\footnote{Strictly speaking, $3H^2$ in \eqref{Hubbleeq} does not contain the energy density of matter fields. However, the error is negligibly small since the matter energy density is much smaller than the inflaton energy density.}  We find that the KK particle energy density $\rho_{KK}$ to the entropy density $s_R$ ratio to be
\begin{align}
  \frac{\rho_{KK}}{s_R}=\frac{3}{4}r_{KK}T_R
\end{align}
which is a conserved quantity under the cosmic expansion. Thus, the condition that the KK particle energy does not exceed the current dark matter energy density is
\begin{align}
    r_{KK}\leq 5.6\times 10^{-21}\left(\frac{10^{11}{\rm GeV}}{T_R}\right)\label{rKKcond}
\end{align}
where we have taken $\Omega_{\rm DM}=0.26$. We note that the above estimation is based on the assumption that the KK particle number density decreases only by the cosmic expansion. Note, however, that the KK particles are coupled to the U(1) gauge field, and the KK particle and anti-particle pairs can annihilate with each other, which would lead to the Boltzmann equation, 
\begin{align}
    \dot{n}_{KK}+3Hn_{KK}=-\langle\sigma v\rangle n_{KK}^2,
\end{align}
where $\langle\sigma v\rangle$ is the averaged cross section evaluated by the momentum distribution~\eqref{analyticKK} and $n_{KK}(t)=\int \frac{d^3\bm k}{(2\pi)^3}N_{n=1,k}^{(0)}(t)$. Nevertheless, when the initial KK particle number density is very small, the annihilation effects would be negligible. 

The constraint~\eqref{rKKcond} is rather stringent, and we find that an illustrative example given in the previous subsections is excluded. As indicated in \eqref{analyticKK}, the KK particle number density is exponentially sensitive to the bulk mass parameter, and the $\mathcal{O}(1)$ change of the parameter reduces the KK particle number density. In appendix~\ref{illustative2}, we have shown another parameter set that can satisfy the above constraint for $T_R\leq 10^{11}$ GeV. Nevertheless, the possibility of the KK particle overproduction should be taken seriously. As we will show in the next section, the KK particle production can take place even if the Wilson line field $\vartheta$ is not an inflaton but just a spectator field.

We discuss the possible generalizations to the models including the standard model, which is necessary to describe the reheating era. Since the 5D theory compactified on $S^1$ cannot have a chiral gauge theory as its 4D effective theory, it is impossible to realize the standard model from the bulk fields. The simplest possibility is to put a 3-brane on which the standard model particles live. As long as the vacuum energy on the brane, namely the tension of the 3-brane, is small enough, the bulk structure on $S^1$ is not much deformed, and our result can straightforwardly be applied.\footnote{Such localization of the standard model on the brane is implicitly assumed e.g. in the higher dimensional inflation model \cite{Anchordoqui:2023etp,Antoniadis:2023sya,Anchordoqui:2024amx,Hirose:2025pzm}.} It is also possible to construct models on $S^1/Z_2$ rather than $S^1$ where the standard model can be on the fixed point $y=0,\pi R$. In such a case, a naive assignment of the orbifold parity of the gauge field $A_\mu(-y)=A_\mu(y)$, $A_5(-y)=-A_5(y)$ removes the zero mode of $A_5$, namely no Wilson line modulus. Therefore, we need to assign ``odd'' parity $A_5(-y)=A_5(y)$ while $A_\mu(-y)=-A_\mu(y)$. Note that unless the vacuum energy on the fixed points (namely, the brane tension) is small, the bulk is flat and our result should hold in the orbifold model, although one needs to take into account the fact that half of the KK modes should be projected out. Another way is to consider a higher-dimensional theory. For instance, the magnetized tori models~\cite{Cremades:2004wa,Abe:2008sx,Abe:2012ya,Abe:2012fj} realize a chiral spectrum in the 4D effective theory with and without orbifold projections. As we demonstrated in~\cite{Abe:2024mka}, the KK-Schwinger effect takes place in 6D cases.\footnote{If matter fields couple to magnetic fluxes, the KK mass levels do not change by an electric field and the KK Schwinger effect does not take place. This means that the KK modes of the standard model sector realized within the magnetized tori models would not be produced. }

Finally, we also comment on the situation where the produced KK particles promptly decay into lighter elements. As the stability of the KK particles is a consequence of the translational invariance of the compact space, which implies momentum conservation, the violation of translational invariance along the compact space would break the stability. As mentioned above, one way to embed the standard model particles is to put a 3-brane in the bulk, which breaks the translational invariance. Even if tree-level couplings between the standard model particles and the KK particles are absent, couplings between them can be generated through gravitational loop corrections, for example. We expect the violation of the translational invariance along the compact space would allow the decay (rather than the co-annihilation) of the KK particles into the standard model particles. As the mass of the KK particles is superheavy, the decay rate is estimated to be large enough that the KK particles would not affect the late time universe. It would be interesting to consider the fate of the KK particles within such a setup, which is beyond the scope of this work.

\section{KK particle production due to misaligned Wilson line modulus}\label{misalignment}
So far, we have considered an illustrative example where the Wilson line modulus $\vartheta$ is an inflaton field. However, KK-Schwinger effect occurs even if the Wilson line modulus $\vartheta$ is not an inflaton but just a spectator field. In this section, we assume that the inflation is driven by another field, and the inflation scale is much larger than the potential scale of $\vartheta$.

\begin{figure}[tbp]
  \begin{minipage}[b]{0.48\columnwidth}
    \centering
    \includegraphics[width=\columnwidth]{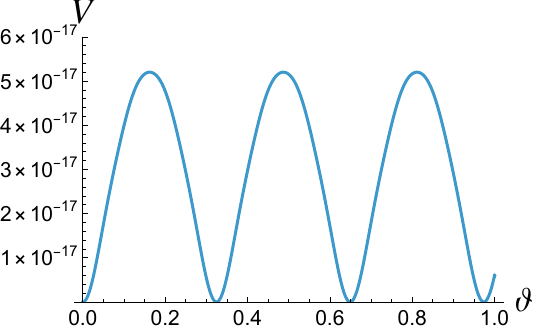}
    \caption{The one-loop effective potential of $\vartheta$ with the parameters~\eqref{parameter2}.}
    \label{fig:potential2}
  \end{minipage}
  \hspace{0.04\columnwidth} 
  \begin{minipage}[b]{0.48\columnwidth}
    \centering
    \includegraphics[width=\columnwidth]{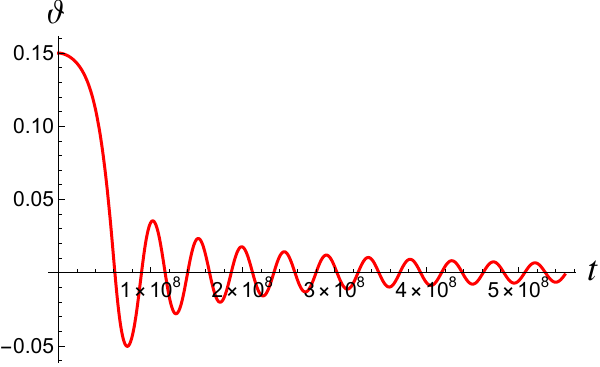}
    \caption{The time evolution of $\vartheta$. The initial condition is $\vartheta=0.15$. The backgroud matter energy is taken to be $\rho_0=10^{-13}$.}
     \label{fig:theta2}
  \end{minipage}
\end{figure}

Let us consider a concrete setup with the parameters
\begin{align}
g_0=&0.1, \quad 2\pi R\Lambda^5=2.83759\times10^{-7},\quad R=40,\nonumber \\
    (N_A,m_{0,A},q_A)=&\ (28,3.8\times10^{-2},0), \quad (2,5\times 10^{-6},1),\quad (2,2\times10^{-4},5),\nonumber\\
    (N_B,m_{1/2,B},q_B)=&\ (16,3.9\times10^{-2},0),\quad (1,6\times10^{-6},1),\quad (1,2\times10^{-4},5),\label{parameter2}
\end{align} 
where we have chosen the hierarchical bulk mass for the neutral sector to make the radion heavier, which is not essential for the following discussion, but just for simplicity. The radion $\varphi$ is stabilized at $\varphi=-0.32$, and the resultant volume of the compact space is
\begin{align}
    L=2\pi R \langle b\rangle\simeq193,
\end{align}
and the effective potential for the Wilson line modulus $\vartheta$ is shown in Fig.~\ref{fig:potential2}. In this case, the effective decay constant is sub-Planckian, and therefore, the subtlety associated with quantum gravity corrections would be reduced.
We assume that inflation is driven by some other field, and as is well known, a light scalar field fluctuates during inflation.\footnote{A scalar field two point function is given by $\langle \Phi^2\rangle=\frac{H^3}{4\pi^2}t$~\cite{Linde:1982uu,Starobinsky:1982ee,Vilenkin:1982wt}, which implies that the scalar field value $\Phi$ jumps over about $H$ per the time $H^{-1}$. So, the ``initial field value'' of $\Phi$ after inflation is stochastically determined.} After inflation, the universe is dominated e.g. by the inflaton energy density described by $\rho_m=\rho_0\left(\frac{a_0}{a(t)}\right)^3$, and the Wilson line modulus is at $\vartheta=\vartheta_0$ at the ``initial time'' $t_0$ ($a(t_0)=a_0$). This is nothing but the standard misalignment mechanism for the axion dark matter scenario. More explicitly, the total energy density is written as
\begin{align}
    \rho_{\rm tot}=\rho_0\left(\frac{a_0}{a(t)}\right)^3+\frac12\dot\vartheta^2+V_{\rm loop}.
\end{align}
As a numerical example, we take $\rho_0=10^{-13}$ and show the resultant $\vartheta$-dynamics in Fig.~\ref{fig:theta2}. As the axion misalignment mechanism, the Wilson line modulus starts to oscillate around the time $H\sim m_\vartheta$ where $m_\theta$ is the mass of the Wilson line modulus. 
As $\vartheta$ rolls down to the bottom, KK Schwinger effect takes place in the same way as the extra-natural inflation, even though the energy scale is much smaller than that case. 

\begin{figure}
    \centering
    \includegraphics[width=0.8\linewidth]{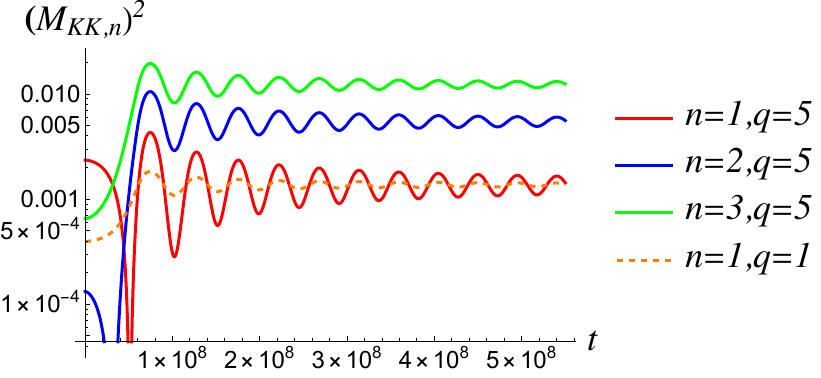}
    \caption{Time dependence of KK masses within the misalignment Wilson line model. The $n=1,2$ modes with $q=5$ experience vanishing of the KK mass, which indicates the KK particle production.  }
    \label{fig:KKmass2}
\end{figure}
\begin{figure}
    \centering
    \includegraphics[width=0.8\linewidth]{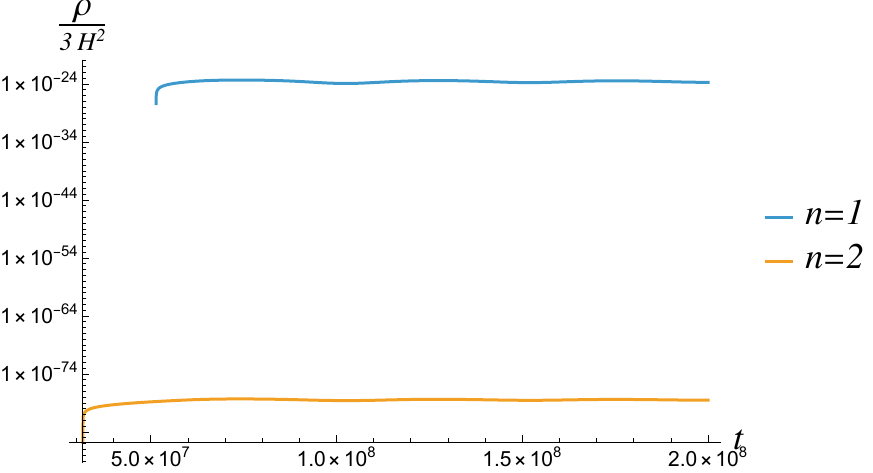}
    \caption{The energy density of the KK particles normalized by the total energy density $3H^2$. Here we have taken the bulk scalar mass to be $m_0=10^{-6}$.}
    \label{fig:ratio2}
\end{figure}

Let us consider the particle production within this setup. We show the time-dependence of the KK mass in Fig.~\ref{fig:KKmass2}, which shows that $n=1,2$ $q=5$ modes can be produced from the vacuum. Recalling that the estimation of the particle production \eqref{analyticKK} does not rely on the scale, we continue to use the formula to estimate physical quantities. We have numerically derived the particle production time from the behavior of the KK masses and derived the values of $\dot\vartheta$ at the production time, which yields the estimation of the energy density of the KK particles shown in Fig.~\ref{fig:ratio2}. This shows that the KK particle energy is yet small compared with the total energy density but still non-negligible. We also note that the KK mass in this model is $M_{KK,n}\sim 0.03 n=7.2n\times10^{16}$ GeV as read from Fig.~\ref{fig:KKmass2}. Thus, superheavy KK particles can be produced in the same way as the extra-natural inflation case. 

Although the relic density of the KK particle in this example is negligible, if the reheating temperature is about $\mathcal{O}(10^{14})$ GeV, the $n=1$ KK particles can explain the whole dark matter today. Notice, however, that we have to choose a very large bulk mass for the $q=5$ sector compared with the one for $q=1$. Otherwise, the KK particles overclose the universe depending on the reheating temperature. In other words, if bulk masses are comparable, it is possible that the KK particles of large charge sectors can be easily produced, which indicates a possible moduli problem within the higher-dimensional theory. 

\section{Conclusion and discussions}\label{conclusion}
In this paper, we have shown that the Wilson line modulus originating from the U(1) gauge potential along the compact space can lead to KK particle production in the early universe, on the basis of the previous work~\cite{Yamada:2024aca,Abe:2024mka}. We have explicitly shown the construction of the extra-natural inflation setup from the 5D theory by directly evaluating the one-loop energy-momentum tensor of scalars and spinors in the time-dependent background spacetime, which was not discussed in our previous work.

The Wilson line modulus acquires an effective potential consisting of sinusoidal terms, and the ``natural inflation'' is realized as shown in the original work~\cite{Arkani-Hamed:2003wrq,Arkani-Hamed:2003xts,Kaplan:2003aj}. In light of the recent CMB observation results, we argue that the multi-natural extension~\cite{Czerny:2014wza,Czerny:2014xja}, which can be realized by adding charged matter fields. As illustrated in our toy model in Sec.~\ref{illustrative}, the potential can be deformed by the additional charged fields, and the inflationary predictions can be consistent with the observational results. In our model, we have just fine-tuned model parameters such as bulk mass. We note that if we add more U(1) symmetries or compact space dimensions, there can appear the ``axion landscape'' considered e.g. in~\cite{Higaki:2014mwa,Higaki:2014pja}. In the landscape of axion potential, there can be (accidentally) an appropriate inflationary direction that explains the CMB observational results.

The introduction of various charged fields leads to the possibility of the KK particle production via the KK-Schwinger effect. In our toy model, we have numerically shown that the KK particles are indeed produced during and after inflation, as expected from the flat spacetime models in \cite{Yamada:2024aca,Abe:2024mka}. Such a mechanism produces the KK particles despite the hierarchy between the KK scale and the inflation scale. This is not surprising as the ``KK modes'' at the final time could be the ``zero mode'' at the time of their production.

We have also discussed another scenario in Sec.~\ref{misalignment} where the Wilson line modulus is not an inflaton but just a spectator field, which is the same as the misalignment mechanism for the axion dark matter scenario. As we have shown explicitly, the KK-Schwinger effect occurs in exactly the same way as the extra-natural inflation case, which shows the generality of the KK-Schwinger effect. This implies that the KK-Schwinger effect is a rather general phenomenon that occurs in the early universe. 

The superheavy KK particles produced from the vacuum significantly affect the reheating dynamics and the present universe if they are stable. As we have described in the toy models, the amount of the KK particles is sensitive to the model parameters. If the KK particles are unstable, the produced KK particles would immediately decay into light species, as in the case of the instant preheating~\cite{Felder:1998vq,Felder:1999pv},\footnote{See also \cite{Taya:2022gzp} for a more precise scattering description in the instant preheating.} and the reheating is completed by the decay of the inflaton. However, at least in our setup, the lowest KK modes $n=\pm1$ are stable since the KK number is a conserved quantity due to the translational invariance of the compact space. Therefore, only the annihilation process reduces the lowest KK particles. As a result of the stability, we have obtained a stringent constraint~\eqref{rKKcond}, which can be applied independently of whether the Wilson line modulus $\vartheta$ is either an inflaton or a spectator field.

Since the KK-Schwinger effect is a rather general mechanism producing KK particles, it would be better to discuss when the phenomenon does NOT occur. We should emphasize again that whether the Wilson line moduli play the role of the inflaton does not matter at all, but whether the moduli move on their field space during or after inflation does. The simplest way to prevent the phenomenon is to remove the zero mode of the gauge fields along the compact spaces. This is achieved e.g. by orbifold projections. For instance, in the 5D case on $S^1/Z_2$ instead of $S^1$, the simple assignment of the orbifold parity for the vector fields is $A_\mu(x,-y)=A_\mu(x,y),\ A_y(x,-y)=-A_y(x,y)$, which is often called an ``even'' parity and forbids the zero mode of $A_y$, and therefore no light gauge field along the compact space. Also, as long as the zero mode of $A_y$, if exists, is heavy and does not move during and after inflation, the KK-Schwinger effect does not occur. Note, however, that there exists an orbifold parity, called an ``odd'' parity, $A_\mu(x,-y)=-A_\mu(x,y),\ A_y(x,-y)=A_y(x,y)$, which leaves the zero mode of $A_y$.\footnote{In such a case, the gauge coupling should also be a singular form so that the covariant derivative properly transforms under orbifold parity transformations. Such a singular coupling is naively incompatible e.g. with supergravity but can be embedded into supergravity~\cite{Bergshoeff:2000zn}. (See also \cite{Bergshoeff:2001pv,Fujita:2001bd,Kugo:2002js}.)} Another way to prevent the KK-Schwinger effect is to modify the KK level structure. Indeed, within 6D models with magnetic fluxes along tori, the KK level structure differs from the case without magnetic fluxes, and it is shown that the time-dependence of the Wilson lines does not change the KK levels thanks to the magnetic fluxes, and as a result, no KK-Schwinger effect occurs~\cite{Abe:2024mka}. Taking account of these observations, it is an interesting question whether the KK particle production occurs within superstring models and how the KK particle production affects the string cosmology, which has not been much discussed as far as we know.

Let us finally discuss future directions. One of the interesting directions is to embed our model into 5D supergravity as a partial UV completion of our models. On the orbifold $S^1/Z_2$, it is possible to introduce the standard models on the branes at the orbifold fixed points, and such model building would be possible on the basis of the supergravity formulation in \cite{Fujita:2001bd,Kugo:2002js}. Also, it is interesting to consider the backreaction effects on the dynamics, which are not discussed in this work. The backreaction during inflation would lead to interesting observational signatures~\cite{Green:2009ds,Flauger:2016idt},\footnote{Indeed, our model can be regarded as a partial UV completion of the model studied in \cite{Flauger:2016idt}, and the analysis performed there can directly be applied.} and the backreaction after inflation would affect the relic abundance of KK particles, which is relevant to the present universe. The backreaction issue generally requires huge computational resources simply because the number of degrees of freedom in quantum field theory is infinite. Another direction of the study is to investigate the models with non-Abelian gauge theories. 10D super-Yang-Mills theory on magnetized tori is a semi-realistic model having the standard-model-like particle spectrum, and the ``matter fields'' are from the off-diagonal components of an adjoint representation of a larger gauge group. Then, one naively thinks that the ``matter fields'' may excite some KK modes, since they are originally ``gauge fields''. If it is the case, the motion of the (supersymmetric) standard-model sector may lead to an analogue of the KK-Schwinger effect. We leave those questions for future work.
\section*{Acknowledgment}
We would like to thank Hiroyuki Abe and Junichiro Kawamura for discussions. YY is supported by IBS under the project code IBS-R018-Y3-2025-a00.
\appendix
\section{Notation}\label{notation}
Throughout this work, we use the convention in \cite{Wess:1992cp} where the 4D Pauli matrices are given by
\begin{align}
    \sigma^0=\left(\begin{array}{cc}
    -1&0\\
    0&-1\end{array}\right),\quad \sigma^1=\left(\begin{array}{cc}
    0&1\\
    1&0\end{array}\right),\quad \sigma^2=\left(\begin{array}{cc}
    0&-\ri\\
    \ri&0\end{array}\right),\quad \sigma^3=\left(\begin{array}{cc}
    1&0\\
    0&-1\end{array}\right),
\end{align}
$\bar{\sigma}^0=\sigma^0$ and $\sigma^i=-\bar{\sigma}^i$ ($i=1,2,3$). Nevertheless, we use the forumulas shown in \cite{Dreiner:2008tw} by appropriately taking into account the differences of the convention.

We also need to specify the convention for 5D spinors, which is the same as 4D Dirac spinors. A 4D Dirac spinor in \cite{Wess:1992cp} can be understood as 
\begin{align}
    \Psi=\left(\begin{array}{c}\chi_\alpha\\ \bar\psi^{\dot\alpha}\end{array}\right),
\end{align}
where $\chi_\alpha$ and $\bar{\psi}^{\dot\alpha}$ are two-component spinors. We define the $\gamma$-matrices as
\begin{align}
    \gamma^\mu=\ri\left(\begin{array}{cc}
      0   &  \sigma^\mu\\
       \bar{\sigma}^\mu  &0 
    \end{array}\right),
\end{align}
by the 4D Pauli matrices $\{\sigma^\mu\}$ defined above. We emphasize that this is different from the $\gamma$-matrix defined in \cite{Wess:1992cp}, and we have defined it so that $\{\gamma^\mu,\gamma^\nu\}=2\eta^{\mu\nu}$. We also define $\gamma^5$ as
\begin{align}
    \gamma^5=\ri\gamma^0\gamma^1\gamma^2\gamma^3=\left(\begin{array}{cc}
       {\bm 1}_{2\times2}  &0  \\
        0 &  -{\bm 1}_{2\times2}
    \end{array}\right)
\end{align}
which again differs from the one in \cite{Wess:1992cp}. With this $\gamma^5$, our $\gamma$-matrices satisfy the 5D Clifford algebra $\{\gamma^M,\gamma^N\}=2\eta^{MN}{\bm 1}_{4\times4}$ ($M,N=0,1,2,3,5$).

\section{Expectation values of the energy-momentum tensor}\label{EMtensor}
We evaluate the expectation value of the energy-momentum tensor operator of various fields. Specifically, we discuss the expectation value of spin 0 and 1/2 fields in the adiabatic vacuum, whereas that of a U(1) vector field is at the classical background level. The energy-momentum tensor of spin 0, 1/2, 1 is written, respectively, as
\begin{align}
    T_{MN}^{(0)}=&2D_M\Phi D_N\bar{\Phi}+g_{MN}\left(-D_P\Phi D^P\bar{\Phi}-m_0^2|\Phi|^2\right),\\
    T_{MN}^{(1/2)}=&-\frac12\bar{\Psi}\gamma_{(M}\overleftrightarrow{D}_N)\Psi+g_{MN}\left(-\bar\Psi\gamma^PD_P\Psi+m_{1/2}\bar\Psi\Psi\right)\nonumber\\
    \overset{\text{on-shell}}{\to}&-\frac12\bar{\Psi}\gamma_{(M}\overleftrightarrow{D}_N)\Psi,\\
    T_{MN}^{(1)}=&F_{MP}F_N{}^{P}-\frac{1}{4}g_{MN}F_{PQ}F^{PQ},
\end{align}
where $A_{(M}B_{N)}=\frac12\left(A_MB_N+A_NB_M\right)$. We note that a bulk cosmological constant term yields $  T^\Lambda_{MN}=-g_{MN}\Lambda^5$.

\subsection{scalar fields}
The energy momentum tensor components having non-vanishing expectation values are only $T_{00}$, $T_{ii}$ $(i=1,2,3)$ and $T_{55}$. We evaluate $\langle T_{00}\rangle$ in detail, and the other components are similarly evaluated. Substituting the expansion~\eqref{0KKexpansion} as well as \eqref{spin0Q} (and their conjugate) yields
\begin{align}
    \langle \hat{T}_{00}\rangle=&\left\langle2\dot{\hat\Phi}\dot{\hat{\overline\Phi}}{}-\frac{1}{b}\left(b\dot{\hat\Phi}\dot{\hat{\Phi}}^\dagger-\frac{b}{a^2}\partial_i\hat\Phi\partial_i\hat{\Phi}^\dagger-b^{-2}D_y\hat\Phi D_y\hat{\Phi}^\dagger-m_0^2\hat{\Phi}\hat{\Phi}^\dagger\right)\right\rangle\nonumber\\
    =&\frac{1}{2\pi R a^3}\sum_n\Biggl\langle\left(\dot{\hat{\phi}}_n-\frac32H_a{\hat{\phi}}_n\right)\left(\dot{{\hat{\phi}}}_n^\dagger-\frac32H_a{\hat{\phi}}_n^\dagger\right)+\partial_i\hat{\phi}_n\partial_i\hat{\phi}^\dagger_n+\frac{1}{b^3}\left(\frac{n}{R}- qg A_5\right)^2\left|\hat{\phi}_n\right|^2+\frac{m_0^2}{b}\left|\hat{\phi}_n\right|^2\Biggr\rangle\nonumber\\
   =&\frac{1}{2\pi R a^3}\int \frac{d^3\bm k}{(2\pi)^3}\sum_n\left[\left|\dot{f}_{n,k}-\frac32H_af_{n,k}\right|^2+\left(\omega_{0,n,k}^2+\frac32\dot{H}_a+\frac94H_a^2\right)|f_{n,k}|^2\right]\nonumber\\
   =&\frac{1}{2\pi R a^3}\int \frac{d^3\bm k}{(2\pi)^3}\sum_n\Biggl[|\dot{f}_{n,k}|^2+\left(\omega_{0,n,k}^2+\frac32\dot{H}_a+\frac92H_a^2\right)|f_{n,k}|^2-\frac{3H_a}{2}\frac{d}{dt}|f_{n,k}|^2\Biggr],
\end{align}
where we have evaluated the quantum operators in the past vacuum states annihilated by $\hat{a}_{n,\bm k},\hat{b}_{n,\bm k}$.\footnote{Nevertheless, the following results are applied to the case e.g. with a future vacuum since the past or future vacuum conditions are implicitly imposed on the asymptotic behavior of $f_{n,k}$.}
Noting that
\begin{align}
    \dot{f}_{n,k}(t)=\frac{\ri\omega_{0,n,k}}{\sqrt{2\omega_{0,n,k}}}\left(\tilde{\alpha}_{n,k}-\tilde{\beta}_{n,k}\right)
\end{align}
we find
\begin{align}
    &\left|\dot{f}_{n,k}\right|^2=\frac{\omega_{0,n,k}}{2}\left(1+2N_{n,k}^{(0)}-2R^{(0)}_{n,k}\right)\\
    &|f_{n,k}|^2=\frac{1}{2\omega_{0,n,k}}\left(1+2N_{n,k}^{(0)}+2R_{n,k}^{(0)}\right),\\
    &\frac{d}{dt}|f_{n,k}|^2=-I_{n,k}^{(0)},
\end{align}
where we have introduced  $\tilde\alpha_{n,k}\overline{\tilde{\beta}}_{n,k}=R^{(0)}_{n,k}+\ri I^{(0)}_{n,k}$ with $R_{n,k},I_{n,k}\in {\mathbb R}$ and $N^{(0)}_{n,k}=|\beta_{n,k}|^2$.
Thus, the energy density $\langle\hat{T}_{00}\rangle$ is
\begin{align}
    \langle\hat{T}_{00}\rangle=&\frac{1}{2\pi R a^3}\sum_{n\in \mathbb Z}\int\frac{d^3\bm k}{(2\pi)^3}\Biggl[\omega_{0,n,k}+\frac34\frac{\dot{H}_a+3H_a^2}{\omega_{0,n,k}}+2\left(\omega_{0,n,k}+\frac{3}{4}\frac{\dot{H}_a+3H_a^2}{\omega_{0,n,k}}\right)N_{n,k}^{(0)}\nonumber\\
   &\hspace{4cm} +\frac32\frac{\dot{H}_a+3H_a^2}{\omega_{0,n,k}}R_{n,k}^{(0)}+\frac{3H_a}{2}I_{n,k}^{(0)}\Biggr].
\end{align}
Notice that, except for the first two terms, all the terms in the square bracket implicitly contain $|\beta_{n,k}|$, which typically decays exponentially in $k$. Then, such terms lead to no UV divergences from the large $k$ integral region, whereas the first two terms can. Therefore, in performing the momentum integration for the first line, we use dimensional regularization, which gives
\begin{align}
    \langle\hat{T}_{00}\rangle_{\rm vac}
    =&\frac{1}{2\pi R a^{D}}\sum_{n\in \mathbb Z}\mu^{3-D}\int\frac{d^D\bm k}{(2\pi)^D}\left(\omega_{0,n,k}+\frac34\frac{\dot{H}_a+3H_a^2}{\omega_{0,n,k}}\right)\nonumber\\
    =&\frac{1}{2\pi R}\sum_{n\in \mathbb Z}\frac{2\mu^{3-D}}{\Gamma\left(\frac{D}{2}\right)}\int\frac{d\tilde{k}}{(4\pi)^{D/2}}\tilde{k}^{D-1}\left(\omega_{0,n,k}+\frac34\frac{\dot{H}_a+3H_a^2}{\omega_{0,n,k}}\right)\nonumber\\
    =&\frac{1}{2\pi R}\sum_{n\in \mathbb Z}\frac{\mu^{3-D}}{(4\pi)^{\frac{D+1}{2}}}\left[-\Gamma\left(-\frac{D+1}{2}\right)M_{0,n}^{D+1}+\frac{3(\dot{H}_a+3H_a^2)}{2}\Gamma\left(-\frac{D-1}{2}\right)M_{0,n}^{D-1}\right].
\end{align}
To perform the KK mode sum, we use the following $\zeta$-function formula~\cite{Elizalde:1995hck},\footnote{Although there appear no $\zeta$-function explicitly, this formula is a special case of the Epstein $\zeta$-function.}
\begin{align}
&\sum_{n=-\infty}^\infty\left[A(n+c)^2+Q\right]^{-s}\nonumber\\
=&\sqrt{\frac{\pi}{A}}\frac{\Gamma(s-1/2)}{\Gamma(s)}Q^{\frac12-s}+\frac{4\pi^s}{\Gamma(s)}A^{-\frac14-\frac s2}Q^{\frac14-\frac s2}\sum_{n=1}^\infty n^{s-\frac12}\cos(2\pi n c)K_{s-\frac12}\left(2\pi n\sqrt{\frac QA}\right),
\end{align}
provided $A>0,c\neq 0,-1,-2,\cdots$. Using the above formula for $s=-D/2\mp1/2$, $A=\frac{1}{R^2b^3}$, $c=-qg_0b\vartheta R$ and $Q=\frac{m_0^2}{b}-\frac94H_a^2-\frac32\dot{H}_a$, we find
\begin{align}
    \sum_{n\in \mathbb Z}M_{0,n}^{D+1}=& \frac{\sqrt{\pi}L}{2\pi}\frac{\Gamma\left(-\frac{D}{2}-1\right)}{\Gamma\left(-\frac{D+1}{2}\right)b^{\frac{D+1}{2}}}M_0^{2+D}\nonumber\\
    &+\frac{2^{\frac D2+2}M_0}{\sqrt\pi\Gamma\left(-\frac{D+1}{2}\right)b^{\frac {D+1}{2}}}\left(\frac{M_0}{L}\right)^{\frac{D}{2}}\sum_{n=1}^{\infty}n^{-1-\frac{D}{2}}\cos\left( n qg_0\vartheta L\right)K_{-1-\frac{D}{2}}\left(  n LM_0\right),\\
    \sum_{n\in \mathbb Z}M_{0,n}^{D-1}=& \frac{\sqrt{\pi}L}{2\pi}\frac{\Gamma\left(-\frac{D}{2}\right)}{\Gamma\left(-\frac{D-1}{2}\right)b^{\frac{D-1}{2}}}M_0^{D}\nonumber\\
    &+\frac{2^{\frac D2+1} L}{\sqrt\pi\Gamma\left(-\frac{D-1}{2}\right)b^{\frac{D-1}{2}}}\left(\frac{M_0}{L}\right)^{\frac{D}{2}}\sum_{n=1}^{\infty}n^{-\frac{D}{2}}\cos\left(  n qg_0\vartheta L\right)K_{-\frac{D}{2}}\left( n LM_0\right),
\end{align}
where we have introduced the physical volume of the compact space
\begin{align}
    L\equiv 2\pi Rb,
\end{align}
and the effective scalar mass
\begin{align}
    M_0^2\equiv m_0^2-b\left(\frac32\dot{H}_a+\frac94H_a^2\right).
\end{align}
Thus, we find
\begin{align}
    &\langle\hat{T}_{00}\rangle_{\rm vac}\nonumber\\
    =&\frac{1}{2\pi R}\frac{\mu^{3-D}}{(4\pi)^{\frac{D+1}{2}}}\Biggl[-\frac{L}{2\sqrt{\pi}}\frac{\Gamma\left(-\frac{D}{2}-1\right)M_0^{2+D}}{b^{\frac{D+1}{2}}}-\frac{2^{\frac D2+2}M_0}{\sqrt\pi b^{\frac {D+1}{2}}}\left(\frac{M_0}{L}\right)^{\frac{D}{2}}\sum_{n=1}^{\infty}n^{-1-\frac{D}{2}}\cos\left( n qg_0\vartheta L\right)K_{-1-\frac{D}{2}}\left(  n LM_0\right)\nonumber\\
    &+\frac{3(\dot{H}_a+3H_a^2)}{2}\Biggl\{\frac{L}{2\sqrt{\pi}}\frac{\Gamma\left(-\frac{D}{2}\right)}{b^{\frac{D-1}{2}}}M_0^{D}+\frac{2^{\frac D2+1} L}{\sqrt\pi b^{\frac{D-1}{2}}}\left(\frac{M_0}{L}\right)^{\frac{D}{2}}\sum_{n=1}^{\infty}n^{-\frac{D}{2}}\cos\left(  n qg_0\vartheta L\right)K_{-\frac{D}{2}}\left( n LM_0\right)\Biggr\}\Biggr]\nonumber\\
    \to&\frac{1}{L^5b}\Biggl[\frac{(M_0L)^{5}}{60\pi^2}-\frac{1}{ 2\pi^2}\sum_{n=1}^{\infty}\cos\left( n qg_0\vartheta L\right)e^{-x_{n,0}}\frac{3+3x_{n,0}+x_{n,0}^2}{n^5}\Biggr]\nonumber\\
    &+\frac{3(\dot{H}_a+3H_a^2)}{2L^3}\Biggl\{\frac{(LM_0)^{3}}{48\pi^2}+\frac{1}{4\pi^2}\sum_{n=1}^{\infty}\cos\left(  n qg_0\vartheta L\right)e^{-x_{n,0}}\frac{1+x_{n,0}}{n^3}\Biggr\}
\end{align}
where $x_{n,0}\equiv  n LM_{0}$. Here we have taken the limit $D\to 3$, which has no singular behavior thanks to the $\zeta$-function regularization applied to the infinite sum of the KK mass tower.\footnote{We do not insist that the UV finiteness is a general behavior of this system and would be regularization dependent since we have included a zero mode, which exactly behaves as a 4D field leading to the usual UV divergences for energy density.}

In the same way, we evaluate the spatial components of the energy momentum tensor,
\begin{align}
\langle\hat{T}_{ii}\rangle
=&\left\langle2\partial_i\hat{\Phi}\partial_i\hat{\Phi}^\dagger+\frac{a^2}{b}\left(b\dot{\hat\Phi}\dot{\hat{\Phi}}^\dagger-\frac{b}{a^2}\sum_{j=1}^3\partial_j\hat{\Phi}\partial_j\hat{\Phi}^\dagger-\frac{1}{b^2}D_y\hat{\Phi}D_y\hat{\Phi}^\dagger-m_0^2\hat\Phi\hat{\Phi}^\dagger\right)\right\rangle\nonumber\\
=&\frac{1}{2\pi R a}\sum_{n\in \mathbb Z}\int \frac{d^3\bm k}{(2\pi)^3}\Biggl[\frac{1}{\omega_{0,n,k}}\left(\frac{k^2}{3a^2}-\frac34\dot{H}_a\right)(1+2N_{n,k}^{(0)})-2\left(\omega_{0,n,k}-\frac{\frac{k^2}{3a^2}-\frac34\dot{H}_a}{\omega_{0,n,k}}\right)R_{n,k}^{(0)}+\frac32H_aI_{n,k}^{(0)}\Biggr],
\end{align}
where the sum over $i$ is not taken on the left-hand side and we have replaced $k_i^2$ by $\frac13k^2$, which can be justified by the rotational invariance of the 3-space. As is the $(00)$-component, terms proportional to $|\beta_{n,k}|$ typically decay exponentially in $k$, and therefore lead to no UV divergences. We evaluate separately the vacuum contribution given by
\begin{align}
  \frac{1}{a^2}\langle\hat{T}_{ii}\rangle_{\rm vac}
 =&\frac{1}{2\pi Ra^3}\sum_{n\in \mathbb Z}\int \frac{d^3\bm k}{(2\pi)^3}\left[\frac{1}{\omega_{0,n,k}}\left(\frac{k^2}{3a^2}-\frac34\dot H_a\right)\right]\nonumber\\
    \to&\frac{1}{L^5b}\Biggl[-\frac{(LM_0)^{5}}{60\pi^2}+\frac{1}{2\pi^{2}}\sum_{n=1}^{\infty}\cos\left( n qg_0\vartheta L\right)e^{-x_{n,0}}\frac{3+3x_{n,0}+x_{n,0}^2}{n^5}\nonumber\\
    &\qquad-\frac{b\dot{H}_aL^2(M_0L)^3}{8\pi}-\frac{3\dot{H}_aL^2}{8\pi^{2}}\sum_{n=1}^{\infty}\cos\left( n qg_0\vartheta L\right)e^{-x_{n,0}}\frac{1+x_{n,0}}{ n^3}\Biggr],
\end{align}
where we have taken the $D\to 3$ limit, which is smooth. 

The 5th component of the energy momentum tensor can be evaluated in the same way, which is given by
\begin{align}
\langle\hat{T}_{55}\rangle=&\left\langle2D_y\hat{\Phi}D_y\hat{\Phi}^\dagger+b^2\left(b\dot{\hat\Phi}\dot{\hat{\Phi}}^\dagger-\frac{b}{a^2}\partial_j\hat{\Phi}\partial_j\hat{\Phi}^\dagger-\frac{1}{b^2}D_y\hat{\Phi}D_y\hat{\Phi}^\dagger-m_0^2\hat\Phi\hat{\Phi}^\dagger\right)\right\rangle\nonumber\\
=&\frac{b^3}{2\pi Ra^3}\sum_{n}\int \frac{d^3\bm k}{(2\pi)^3}\Biggl[-R\frac{\partial}{\partial R}\omega_{0,n,k}-\frac{3\dot H_a}{4\omega_{0,n,k}}+\frac{2}{\omega_{0,n,k}}\left(m_{KK,n}^2-\frac34\dot{H}_a\right)N_{n,k}^{(0)}\nonumber\\
&\hspace{3.2cm}+\frac{1}{\omega_{0,n,k}}\left(\frac{k^2}{a^2}+\frac{m_0^2}{b}+\frac32\dot{H}_a-\frac94H_a^2\right)R_{n,k}^{(0)}+\frac32H_aI_{n,k}^{(0)}\Biggr],
\end{align}
where we have used a formal relation $ \frac{\left(n-\frac{qg_0\vartheta L}{2\pi}\right)^2}{R^2b^3\omega_{0,n,k}}=-R\frac{\partial}{\partial R}\omega_{0,n,k}$.\footnote{Here, we mean that we differentiate $R$ that are not multiplied by $\vartheta$, which is the reason why we used $L$ instead of $2\pi Rb$ in the numerator of the left-hand side. We may perform the infinite sum over $n$ directly, but the result is the same, as one can explicitly check.}
The vacuum contribution of $\langle\hat{T}_{55}\rangle$ is
\begin{align}
  \frac{1}{b^3}\langle\hat{T}_{55}\rangle_{\rm vac}
=&\frac{1}{2\pi Ra^3}\sum_{n\in \mathbb Z}\int\frac{d^3\bm k}{(2\pi)^3}\Biggl[-R\frac{\partial}{\partial R}\omega_{0,n,k}-\frac{3\dot H_a}{4\omega_{0,n,k}}\Biggr]\nonumber\\
\to&-\frac{1}{L^5b}\Biggl[\frac{(M_0L)^{5}}{60\pi^2}+\frac{1}{2\pi^2}\sum_{n=1}^\infty\cos\left(nqg_0\vartheta L\right)e^{-x_{n,0}}\frac{12+12x_{n,0}+5x_{n,0}^2+x_{n,0}^3}{n^5}\Biggr]\nonumber\\
&-\frac{3\dot{H}_a}{2L^3}\Biggl\{\frac{(LM_0)^{3}}{48\pi^2}+\frac{1}{4\pi^2}\sum_{n=1}^{\infty}n^{-\frac{3}{2}}\cos\left(  n qg_0\vartheta L\right)e^{-x_{n,0}}\frac{1+x_{n,0}}{n^3}\Biggr\},
\end{align}
where we have taken the limit $D\to 3$ in the second line. 
\subsection{spinor fields}
Let us evaluate the energy-momentum tensor operators of spinor fields in the adiabatic vacuum. The on-shell version of the energy-momentum tensor operator is 
\begin{align}
\hat{T}_{MN}^{(1/2)}=\frac12\hat{\overline{\Psi}}\gamma_{(M}\overleftrightarrow{D}_N)\hat{\Psi}.
\end{align}
In particular, we are interested only in the diagonal components, which would give non-vanishing expectation values. 

The time-component $\langle \hat{T}^{(1/2)}_{00}\rangle$ is rather simply given by \footnote{One should notice that $\gamma_0\partial_0=-\gamma^0\partial_0$.}
\begin{align}
    \left\langle \hat{T}^{(1/2)}_{00}\right\rangle=&\frac12\left\langle\hat{\bar{\Psi}}\gamma_{0}\dot{\hat{\Psi}}-\dot{\hat{\bar{\Psi}}}\gamma_0\hat{\Psi}\right\rangle\nonumber\\
   =&\frac{1}{2\pi Ra^3}\sum_{n\in \mathbb Z,h=\pm}\int \frac{d^3\bm k}{(2\pi)^3}\Biggl[-\omega_{n,k}+2\omega_{n,k}N^{(1/2)}_{n,k,h}+2\mu_{n}\left(\frac{\omega_{n,k}-\frac{kh}{a}}{\omega_{n,k}}\right)R_{n,k,h}^{(1/2)}\Biggr],
\end{align}
where $N^{(1/2)}_{n,k,h}=|\delta_{n,k,h}|^2$ is the particle number density in phase space of spin 1/2 particles and $R_{n,k,h}={\rm Re}\left(\tilde{\gamma}_{n,k,h}\bar{\tilde{\delta}}_{n,k,h}\right)$. Notice that there appear no Hubble-induced terms proportional to $H_a$, namely the curvatures. As in the scalar field case, we separate the vacuum fluctuation and the particle energy density part, and the latter is given by
\begin{align}
\left\langle\hat{T}_{00}^{(1/2)}\right\rangle_{\rm vac}&=-\frac{1}{\pi Ra^3}\sum_{n\in \mathbb Z}\int \frac{d^3\bm k}{(2\pi)^3}\omega_{n,k}\nonumber\\
    &=\frac{1}{L^5b}\Biggl[-\frac{(Lm_{1/2})^{5}}{30\pi^2}+\frac{1}{\pi^{2}}\sum_{n=1}^{\infty}\cos\left( n qg_0\vartheta L\right)e^{-x_{n,1/2}}\frac{3+3x_{n,1/2}+x_{n,1/2}^2}{n^5}\Biggr],
\end{align}
where $x_{n,1/2}\equiv  n Lm_{1/2}$ and note that the one-particle energy is independent of the helicity, which leads to the first equality. It would be worth clarifying the relation to that of a complex scalar field. Compared to a complex scalar field, the vacuum fluctuation contribution (the first term in the integrand) is twice as large as that of a complex scalar, in addition to the overall sign difference. This is because the 5D spinor is decomposed into two 4D Weyl spinors whose degrees of freedom are the same as those of a complex scalar. 
For the 3-space components $\langle T_{ii}\rangle$, the rotational symmetry of 3-spaces implies
\begin{align}
\left\langle\hat{T}_{ii}^{(1/2)}\right\rangle=&\frac16\sum_{j=1}^3\left\langle\bar{\Psi}{\gamma_{j}}\overleftrightarrow{ D}_j\Psi\right\rangle\qquad\left(\text{averaged over the spatial index}\right)\nonumber\\
=&\frac{1}{3(2\pi R)a^2}\sum_{n\in \mathbb Z,h=\pm}\int \frac{d^3\bm k}{(2\pi)^3}\Biggl[-\frac{k^2}{a\omega_{n,k}}\left(1-2N_{n,k,h}^{(1/2)}\right)+\frac{2hk\mu_{n}}{\omega_{n,k}}R_{n,k,h}^{(1/2)}\Biggr].
\end{align}
The vacuum contribution is
\begin{align}
  & \frac{1}{a^2} \left\langle\hat{T}_{ii}^{(1/2)}\right\rangle_{\rm vac}\nonumber\\
 & =-\frac{1}{3(2\pi R)a^3}\sum_{n\in \mathbb Z}\int \frac{d^3\bm k}{(2\pi)^3}\frac{k^2}{a^2\omega_{n,k}}\nonumber\\
   &=\frac{1}{L^5b}\Biggl[\frac{(Lm_{1/2})^{5}}{30\pi^2}-\frac{1}{\pi^{2}}\sum_{n=1}^{\infty}\cos\left( n qg_0\vartheta L\right)e^{-x_{n,1/2}}\frac{3+3x_{n,1/2}+x_{n,1/2}^2}{n^5}\Biggr].
\end{align}
Finally, we consider the 5th component of the energy-momentum tensor given by
\begin{align}  \left\langle\hat{T}_{55}^{(1/2)}\right\rangle=&\frac{b}{2}\left\langle\hat{\overline\Psi}\gamma^{\mathtt 5} \overleftrightarrow{D}_5\hat\Psi\right\rangle\nonumber\\
 &=\frac{b^3}{2\pi R a^3}\sum_{n\in \mathbb Z,h=\pm}\int\frac{d^3\bm k}{(2\pi)^3}\Biggl[-\frac{M_{KK,n}^2}{\omega_{n,k}}\left(1-2N^{(1/2)}_{n,k,h}\right)+\frac{2M_{KK,n}^2kh}{a\omega_{n,k}\mu_n }R^{(1/2)}_{n,k,h}\nonumber\\
 &\hspace{5cm} -\frac{2\left(\frac{n}{R}-qg_0 b\vartheta\right)m_{1/2}}{\mu_nb^2}I_{n,k,h}^{(1/2)}\Biggr].
\end{align}
In particular, we find the vacuum contribution $\langle\hat{T}_{55}\rangle$ as
\begin{align}
    &\frac{1}{b^3}\left\langle\hat{T}^{(1/2)}_{55}\right\rangle_{\rm vac}\nonumber\\
    &=-\frac{1}{2\pi R a^3}\sum_{n\in \mathbb Z}\int\frac{d^3\bm k}{(2\pi)^3}\left(n-\frac{qg_0 \vartheta L}{2\pi} \right)^2\frac{2}{R^2b^3\omega_{n,k}}\nonumber\\
    &=\frac{1}{L^5b}\Biggl[\frac{(Lm_{1/2})^5}{30\pi^2}+\frac{1}{\pi^2}\sum_{n=1}^\infty\cos(n q g_0b\vartheta L )e^{-x_{n,1/2}}\frac{12+12x_{n,1/2}+5x_{n,1/2}^2+x_{n,1/2}^3}{n^5}\Biggr].
\end{align}
\subsection{vector fields}
We derive the energy-momentum tensor due to an Abelian vector field. Let us first consider the contribution from a background field. We take $A_5(t)$ as a background field while keeping others turned off, which yields
\begin{align}
    T_{00}^{(1)}=&F_{0M}g^{MN}F_{0N}-\frac14g_{00}F_{MN}F^{MN}=\frac{1}{2b^2}(\dot{A}_5)^2=\frac{1}{4\pi R}\left(\dot\vartheta+H_b\vartheta\right)^2,\\
    T_{ii}^{(1)}=&-\frac{a^2}{2b^2}(\dot{A}_5)^2=-\frac{1}{4\pi R}\left(\dot\vartheta+H_b\vartheta\right)^2,\\
    T_{55}^{(1)}=&\frac{b}{2}(\dot{A}_5)^2=\frac{b^3}{4\pi R}\left(\dot\vartheta+H_b\vartheta\right)^2.
\end{align}

\section{Complete form of the equations of motion}\label{completeEOM}
In this section, we summarize the complete form of the equations of motion without dropping any terms. The Einstein equation reads
\begin{align}
    3H_a^2-\frac34H_b^2=&\frac{2\pi R}{M_{\rm pl}^2}\rho,\\
    -3H_a^2-2\dot H_a-\frac34H_b^2=&\frac{2\pi R}{M_{\rm pl}^2}p,\\
 -\dot{H}_a+\frac12\dot{H}_b-2H_a^2+\frac32H_a H_b-\frac14H_b^2=&\frac{2\pi R }{3M_{\rm pl}^2}p_y
\end{align}
recalling that $2\pi R M_5^3=M_{\rm pl}^2$ and we have defined $\rho=\langle T_{00}\rangle/a^2,$ $p=\langle T_{ii}\rangle/a^2$ (no sum over $i$) and $p_y=\langle T_{55}\rangle/b^3$. Taking account of contributions from possible scalar and spinor species, we find 
\begin{align}
   2\pi R \rho&=V_{\rm loop}+\sum_{A}\Biggl[\frac{3(\dot{H}_a+3H_a^2)}{2L^3}\Biggl\{\frac{(LM_{0,A})^{3}}{48\pi^2}+\frac{1}{4\pi^{2}}\sum_{n=1}^{\infty}\cos\left( n q g_0\vartheta L\right)e^{-x_{n,0,A}}\frac{1+x_{n,0,A}}{ n^3}\Biggr\}\nonumber\\
    &+\sum_{n\in \mathbb Z}\int\frac{d^3\bm k}{(2\pi)^3a^3}\Biggl\{2\left(\omega_{0,n,k,A}+\frac34\frac{\dot H_a+3H_a^2}{\omega_{0,n,k,A}}\right)N^{(0)}_{n,k,A}+\frac{3H_a}{2}I^{(0)}_{n,k,A}+\frac32\frac{\dot{H}_a+3H_a^2}{\omega_{0,n,k,A}}R^{(0)}_{n,k,A}\Biggr\}\Biggr]\nonumber\\
    &+\sum_{B}\sum_{n\in \mathbb Z,h=\pm}\int \frac{d^3\bm k}{(2\pi)^3a^3}\Biggl\{2\omega_{n,k,B}N^{(1/2)}_{n,k,h,B}+2\mu_{n,B}\left(\frac{\omega_{n,k,B}-\frac{kh}{a}}{\omega_{n,k,B}}\right)R^{(1/2)}_{n,k,h,B}\Biggr\},
\end{align}
where $A,B$ are scalar and spinor species labels with their U(1) charge $q_{A,B}$ and their bulk mass $m_{0,A}$ and $m_{1/2,B}$, respectively. 
The pressure in the 3-space is
\begin{align}
  2\pi R  p=&-V_{\rm loop}+\sum_{A}\sum_{n\in \mathbb Z}\int \frac{d^3\bm k}{(2\pi)^3a^3}\Biggl\{\frac{2}{\omega_{0,n,k,A}}\left(\frac{k^2}{3a^2}-\frac34\dot{H}_a\right)N^{(0)}_{n,k,A}-2\left(\omega_{0,n,k,A}-\frac{\frac{k^2}{3a^2}-\frac34\dot{H}_a}{\omega_{0,n,k,A}}\right)R^{(0)}_{n,k,A}\nonumber\\
  &+\frac{3}{2}H_aI^{(0)}_{n,k,A}\Biggr\}+a\sum_B\sum_{n\in\mathbb Z,h=\pm}\int \frac{d^3\bm k}{(2\pi)^3a^3}\Biggl\{\frac{2k^2}{3a^2\omega_{n,k,B}}N^{(1/2)}_{n,k,h,B}+\frac{2hk\mu_{n,B}}{3a\omega_{n,k,B}}R^{(1/2)}_{n,k,h,B}\Biggr\}.
\end{align}
The pressure along the compact dimension is
\begin{align}
   2\pi R p_y=&\sum_A\Biggl[-\frac{1}{L^5b}\Biggl\{\frac{(M_{0,A}L)^{5}}{60\pi^2}+\frac{1}{2\pi^2}\sum_{n=1}^\infty\cos\left(nq_Ag_0\vartheta L\right)e^{-x_{n,0,A}}\frac{12+12x_{n,0,A}+5x_{n,0,A}^2+x_{n,0,A}^3}{n^5}\Biggr\}\nonumber\\
&\hspace{2cm}-\frac{3\dot{H}_a}{2L^3}\Biggl\{\frac{(LM_{0,A})^{3}}{48\pi^2}+\frac{1}{4\pi^2}\sum_{n=1}^{\infty}\cos\left(  n q_Ag_0\vartheta L\right)e^{-x_{n,0,A}}\frac{1+x_{n,0,A}}{n^3}\Biggr\}\nonumber\\
   &+\sum_{n\in\mathbb Z}\int\frac{d^3\bm k}{(2\pi)^3a^3}\Biggl\{\frac{2}{\omega_{0,n,k,A}}\left(M_{KK,n,A}^2-\frac34\dot{H}_a\right)N_{n,k,A}^{(0)}\nonumber\\
    &+\frac{1}{\omega_{0,n,k,A}}\left(\frac{k^2}{a^2}+\frac{m_{0,A}^2}{b}+\frac32\dot{H}_a-\frac94H_a^2\right)R_{n,k,A}^{(0)}+\frac32H_aI_{n,k,A}^{(0)}\Biggr\}\Biggr]\nonumber\\
    &+\sum_B\Biggl[\frac{1}{L^5b}\Biggl\{\frac{(m_{1/2,B}L)^5}{30\pi^2}+\frac{1}{\pi^2}\sum_{n=1}^\infty\cos(nq_B g_0\vartheta L\theta)e^{-x_{n,1/2,B}}\frac{12+12x_{n,1/2,B}+5x_{n,1/2,B}^2+x_{n,1/2,B}^3}{n^5}\Biggr\} \nonumber\\
  &+\sum_{n\in\mathbb Z,h=\pm}\int\frac{d^3\bm k}{(2\pi)^3a^3}\Biggl\{\frac{2M_{KK,n,B}^2}{\omega_{n,k}}N^{(1/2)}_{n,k,h}-\frac{2M_{KK,n,B}^2}{\mu_{n,B}}\frac{kh}{\omega_{n,k,B}}R^{(1/2)}_{n,k,h,B}\nonumber\\
    &\hspace{4.5cm}+\frac{\left(\frac nR-q_Bg_0b\vartheta\right)m_{1/2,B}}{b^2\mu_{n,B}}I^{(1/2)}_{n,k,h,B}\Biggr\}\Biggr].
\end{align}
Recalling $
    b=\exp\left(\sqrt{\frac{2}{3}}\frac{\varphi}{M_{\rm pl}}\right)$, we find
A linear combination of the Einstein equation leads to the equation of motion of the canonical radion field as
\begin{align}
\ddot\varphi+3H_a\dot\varphi=\sqrt{\frac23}\frac{2\pi R}{M_{\rm pl}}\left(p_y-\frac32p+\frac12\rho\right).\label{radEOM}
\end{align}
we define the right-hand side of \eqref{radEOM} as 
\begin{align}
    \Delta_\varphi\equiv \sqrt{\frac23}\frac{2\pi R}{M_{\rm pl}}\left(p_y-\frac32p+\frac12\rho\right).
\end{align}
The vacuum contribution of $\Delta_\varphi$ is
\begin{align}
    \Delta_\varphi^{\rm vac}\equiv& \sqrt{\frac23}\frac{2\pi R}{M_{\rm pl}}\left(p_y^{\rm vac}-\frac32 p^{\rm vac}+\frac12\rho^{\rm vac}\right)\nonumber\\
 =&
   -\partial_\varphi V_{\rm loop}+\frac{\sqrt6(6 H_a^2+\dot{H}_a)}{M_{\rm pl}bL^2}\sum_A\Biggl[\frac{(LM_{0,A})^{3}}{48\pi^2}+\frac{1}{8\pi^{2}}\sum_{n=1}^{\infty}\cos\left( n q_A g_0\vartheta L\theta\right)e^{-x_{n,0,A}}\frac{1+x_{n,0,A}}{ n^3}\Biggr],
\end{align}
which obviously implies
\begin{align}
    \Delta_\varphi^{\rm vac}\Biggr|_{H_a=0}=-\partial_\varphi V_{\rm loop}
\end{align}
as expected. 
The particle contributions to $\Delta_\varphi$ are 
\begin{align}
    \Delta_\varphi^{\rm real}=&\sqrt{\frac23}\frac{1}{a^3 M_{\rm pl}}\sum_{n\in \mathbb Z}\int\frac{ d^3\bm k}{(2\pi)^3}\Biggl[\sum_{A}\Biggl\{\frac{2}{\omega_{0,n,k,A}}\Biggl\{\left(\frac{m_{0,A}^2}{2b}+\frac{9H_a^2+6\dot{H}_a}{4}+\frac32M_{KK,n,A}^2\right)N^{(0)}_{n,k,A}\nonumber\\
    &+\left(\frac{27{H}_a^2-9\dot{H}_0}{8}+\frac32M_{KK,n,A}^2+\frac{m_{0,A}^2}{2b}-\frac{k^2}{a^2}\right)R^{(0)}_{n,k,A}\Biggr\}\nonumber\\
   & +\sum_B\sum_{h=\pm}\Biggl\{\frac{1}{\omega_{n,k,B}}\left(3M_{KK,n,B}^2+m_{1/2,B}^2b^{-1}\right)N^{(1/2)}_{n,k,h,B}+\frac{2\left(\frac{n}{R}-q_Bg_0 b\vartheta\right)m_{1/2,B}}{b^2\mu_{n,B}}I^{(1/2)}_{n,k,h,B}\nonumber\\
    &+\left(\mu_{n,B}-\frac{4M_{KK,n,B}^2+2m_{1/2,B}^2b^{-1}}{\mu_{n,B}}\frac{kh}{a\omega_{n,k,B}}\right)R^{(1/2)}_{n,k,h,B}\Biggr\}\Biggr].
\end{align}

Let us focus on the dynamical equation of the Wilson line mode $\vartheta$. The background Wilson line modulus follows the equation of motion
\begin{align}
 \ddot\vartheta+3H_a\dot\vartheta-\left(3H_aH_b+\dot H_b+H_b^2\right)\vartheta+(J_0+J_{1/2})=0
\end{align}
where we have defined
\begin{align}
  J_0\equiv& \frac{1}{a^3}\sum_{A}\sum_{n\in \mathbb Z}\partial_\vartheta M_{0,n,A}^2\langle \hat{{\phi}}^A_n\hat{{\phi}}{}^{A\dagger}_n\rangle,\\
  J_{1/2}\equiv&-\frac{1}{a^3}\sum_B\sum_{n\in \mathbb Z}\left(\partial_\vartheta M_{1/2,n}\langle\hat{\zeta}^B_n\hat{\chi}^B_n\rangle+{\rm h.c.}\right).
\end{align}
Each contribution can be evaluated as
\begin{align}
J_0
\to &\sum_A\Biggl[\frac{1}{L^3b^2}\sum_{n=1}^\infty\frac{q_Ag_0}{2\pi^2}\sin(nq_Ag_0\vartheta L)e^{-x_{n,0,A}}\frac{3+3x_{n,0,A}+x_{n,0,A}^2}{n^4}\nonumber\\
&\hspace{2cm}-\frac{2q_Ag_0}{a^3}\sum_{n\in\mathbb Z}\int\frac{d^3\bm k}{(2\pi)^3}\frac{\left(\frac{n}{R}-q_Ag_0b\vartheta\right)}{b^2\omega_{n,k,A}}\left(N_{n,k,A}^{(0)}+R^{(0)}_{n,k,A}\right)\Biggr],\\
J_{1/2}
\to&\sum_B\Biggl[-\frac{1}{L^3b^2}\sum_{n=1}^\infty\frac{q_Bg_0}{\pi^2}\sin(nq_Bg_0\vartheta L)e^{-x_{n,1/2,B}}\frac{3+3x_{n,1/2,B}+x_{n,1/2,B}^2}{n^4}\nonumber\\
&-\frac{2q_Bg_0}{a^3}\sum_{n\in\mathbb Z,h=\pm}\int\frac{d^3\bm k}{(2\pi)^3}\Biggl\{\frac{\left(\frac n R-q_Bg_0b\vartheta\right)}{b^{2}\omega_{n,k,B}}N_{n,k,B}^{(1/2)}+\frac{m_{1/2,B}}{\mu_{n,B}b}I_{n,k,B}^{(1/2)}-\frac{\left(\frac{n}{R}-q_Bg_0\vartheta b\right)}{b^{2}\omega_{n,k,B}}\frac{kh}{a\mu_{n,B}}R_{n,k,B}^{(1/2)}\Biggr\}\Biggr].
\end{align}
One can easily confirm that the sum of the first term of $J_0$ and that of $J_1$ can be written as $\partial_\vartheta V_{\rm loop}$ as expected.

\section{Gradient energy of massless scalar fields and four-form fluxes}\label{gradE}
In the main text, we realize an effective 4D positive cosmological constant by combining the one-loop vacuum energy of matter fields and a 5D cosmological constant. Although not necessary, we give a few comments for a possible alternative to the bulk cosmological constant term. Consider a real massless scalar field equation
\begin{align}
    \partial_M\left(\sqrt{-G}G^{MN}\partial_N\chi\right)=0
\end{align}
which has a solution
\begin{align}
    \partial_y\chi=v\ (={\rm const.})
\end{align}
We may consider such a background configuration as a classical background. The energy-momentum tensor of the real scalar in such a background is
\begin{align}
    T_{00}^\chi=\frac{v^2}{2b^3}=-\frac{1}{a^2}T_{ii}^{\chi}=\frac{1}{b^3}T^{\chi}_{55}.
\end{align}
Such a gradient energy has been considered in \cite{Arkani-Hamed:1999lsd,Anchordoqui:2023etp}. The gradient energy contributes to the potential for the radion field. Note however, the gradient energy is consistent only for an exactly massless scalar, which implies the global shift symmetry for $\chi$. One may wonder if such a global symmetry can be protected, and it seems unlikely that the quantum gravity corrections respect the symmetry.

In the light of such an observation, we give another interpretation of the gradient energy: Let us consider a 3-form field $B_{MNP}$ and its field strength four form $H_{MNPQ}$. We consider a four form flux background along the non-compact direction,
\begin{align}
    H_{0123}=f\ (={\rm const.}) \qquad H_{MNPQ}=0 \ ({\rm otherwise}),
\end{align}
where the anti-symmetry of the indices should be understood. The 4-form has non-vanishing flux along the non-compact direction, and therefore, we do not have a quantization condition on the flux. The dual of the 4-form in 5D is a gauge invariance 1-form $C_M=\epsilon_{MNPQR}H^{NPQR}$, and the 1-form field strength can be written by the 0-form gauge potential  $\tilde{\chi}$ as $C_M=\partial_M\tilde{\chi}$. The flux background implies $C_5\neq 0$, which is nothing but the scalar gradient energy. Thus, we may interpret the gradient energy of a real massless scalar field as the 4-form flux background.

\section{A model with heavier bulk masses}\label{illustative2}
\begin{figure}[tbp]
  \begin{minipage}[b]{0.48\columnwidth}
    \centering
    \includegraphics[width=\columnwidth]{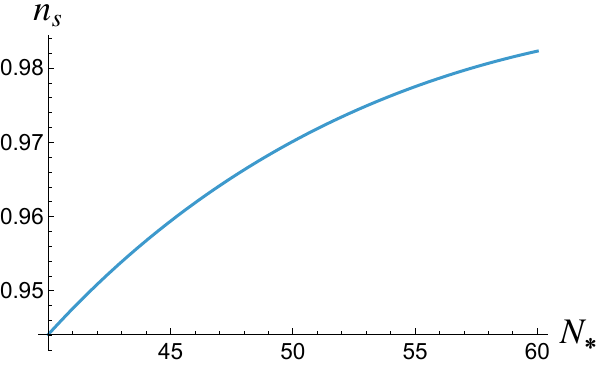}
    \caption{The scalar spectral index $n_s$ as the function of the corresponding e-foloding number $N_*$ of the horizon exit.}
    \label{fig:ns2}
  \end{minipage}
  \hspace{0.04\columnwidth} 
  \begin{minipage}[b]{0.48\columnwidth}
    \centering
    \includegraphics[width=\columnwidth]{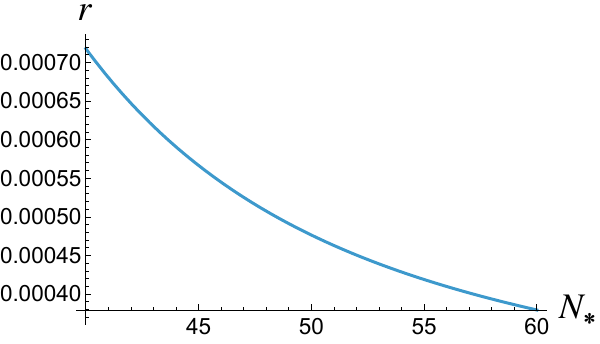}
    \caption{The tensor-to-scalar ratio $r$ as the function of the corresponding e-folding number $N_*$ of the horizon exit.}
     \label{fig:r2}
  \end{minipage}
\end{figure}

\begin{figure}
    \centering
    \includegraphics[width=0.7\linewidth]{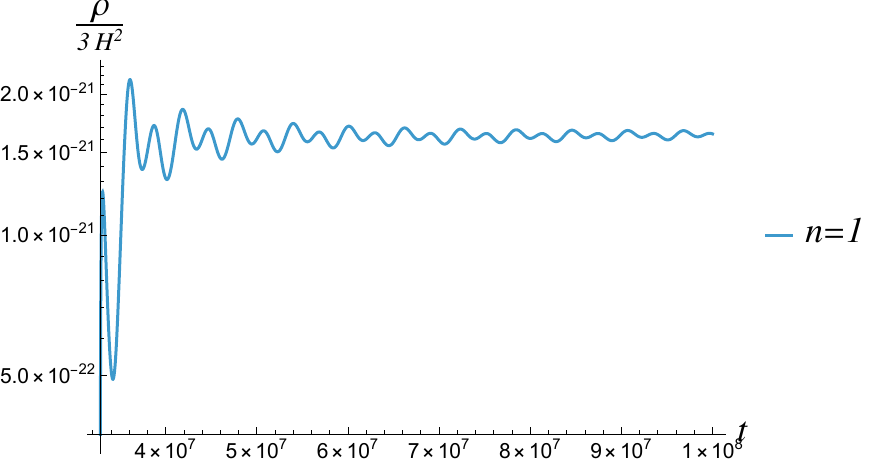}
    \caption{The ratio between the energy density of the KK particle and the Hubble parameter for the case with parameters in \eqref{new parameter}. }
    \label{fig:energyratio2}
\end{figure}

We give another parameter set for the extra-natural inflation where the large charge fields have a larger bulk mass than the one in Sec.~\ref{illustrative}. Here we take the parameters
\begin{align}
    R=&40\times C^{-1}, \ g_0=0.02C,\nonumber\\
    2\pi R\Lambda^5=&1.1182\times 10^{-3}C^4,\nonumber
\end{align}
\begin{align}
    (N_A,m_{0,A},q_A)=&(28,1.8\times10^{-1}C,0), \quad (2,5\times10^{-4}C,1)\quad (2,8.403\times10^{-4}C,5)\nonumber\\
    (N_B,m_{1/2,B},q_B)=&(16,1.9\times 10^{-1}C,0),\quad (1,5.2\times10^{-4}C,1),\quad (1,8.4\times10^{-4}C,5),\label{new parameter}
\end{align}
with the rescaling parameter $C=0.45$. Notice that except for the rescaling parameter, only the bulk masses of the fields with $q=5$ are four times larger than the ones in Sec.~\ref{illustrative}. The resultant potential looks almost the same as the one in Figs.~\ref{fig:3D potential}, \ref{fig:1D potential}. The spectral index and the tensor-to-scalar ratio are slightly different, which are shown in Figs.~\ref{fig:ns2}, \ref{fig:r2}. The KK particle with $n=1,q=5$ is produced, but the amount is much less than the case of Sec.~\ref{illustrative} as shown in Fig.~\ref{fig:energyratio2}. In this case, the ratio can satisfy the constraint in \eqref{rKKcond} for $T_R\leq 10^{11}$ GeV. 

\bibliographystyle{JHEP}
\bibliography{main.bib}

\providecommand{\href}[2]{#2}\begingroup\raggedright\begin{thebibliography}{10}

\bibitem{Guth:1980zm}
A.~H. Guth, {\it {The Inflationary Universe: A Possible Solution to the Horizon
  and Flatness Problems}},  {\em Phys. Rev. D} {\bf 23} (1981) 347--356.

\bibitem{Starobinsky:1980te}
A.~A. Starobinsky, {\it {A New Type of Isotropic Cosmological Models Without
  Singularity}},  {\em Phys. Lett. B} {\bf 91} (1980) 99--102.

\bibitem{Sato:1981qmu}
K.~Sato, {\it {First-order phase transition of a vacuum and the expansion of
  the Universe}},  {\em Mon. Not. Roy. Astron. Soc.} {\bf 195} (1981), no.~3
  467--479.

\bibitem{Linde:1981mu}
A.~D. Linde, {\it {A New Inflationary Universe Scenario: A Possible Solution of
  the Horizon, Flatness, Homogeneity, Isotropy and Primordial Monopole
  Problems}},  {\em Phys. Lett. B} {\bf 108} (1982) 389--393.

\bibitem{Albrecht:1982wi}
A.~Albrecht and P.~J. Steinhardt, {\it {Cosmology for Grand Unified Theories
  with Radiatively Induced Symmetry Breaking}},  {\em Phys. Rev. Lett.} {\bf
  48} (1982) 1220--1223.

\bibitem{Arkani-Hamed:2003xts}
N.~Arkani-Hamed, H.-C. Cheng, P.~Creminelli, and L.~Randall, {\it {Extra
  natural inflation}},  {\em Phys. Rev. Lett.} {\bf 90} (2003) 221302,
  [\href{http://arxiv.org/abs/hep-th/0301218}{{\tt hep-th/0301218}}].

\bibitem{Arkani-Hamed:2003wrq}
N.~Arkani-Hamed, H.-C. Cheng, P.~Creminelli, and L.~Randall, {\it
  {Pseudonatural inflation}},  {\em JCAP} {\bf 07} (2003) 003,
  [\href{http://arxiv.org/abs/hep-th/0302034}{{\tt hep-th/0302034}}].

\bibitem{Kaplan:2003aj}
D.~E. Kaplan and N.~J. Weiner, {\it {Little inflatons and gauge inflation}},
  {\em JCAP} {\bf 02} (2004) 005,
  [\href{http://arxiv.org/abs/hep-ph/0302014}{{\tt hep-ph/0302014}}].

\bibitem{Yamada:2024aca}
Y.~Yamada, {\it {Kaluza-Klein Schwinger effect}},
  \href{http://arxiv.org/abs/2403.13451}{{\tt arXiv:2403.13451}}.

\bibitem{Abe:2024mka}
H.~Abe and Y.~Yamada, {\it {Roles of electric field/time-dependent Wilson line
  in toroidal compactification with or without magnetic fluxes}},  {\em JHEP}
  {\bf 10} (2024) 050, [\href{http://arxiv.org/abs/2408.08517}{{\tt
  arXiv:2408.08517}}].

\bibitem{Furuuchi:2015foh}
K.~Furuuchi, {\it {Excursions through KK modes}},  {\em JCAP} {\bf 07} (2016)
  008, [\href{http://arxiv.org/abs/1512.04684}{{\tt arXiv:1512.04684}}].

\bibitem{Sauter:1931zz}
F.~Sauter, {\it {Uber das Verhalten eines Elektrons im homogenen elektrischen
  Feld nach der relativistischen Theorie Diracs}},  {\em Z. Phys.} {\bf 69}
  (1931) 742--764.

\bibitem{Schwinger:1951nm}
J.~S. Schwinger, {\it {On gauge invariance and vacuum polarization}},  {\em
  Phys. Rev.} {\bf 82} (1951) 664--679.

\bibitem{Czerny:2014wza}
M.~Czerny and F.~Takahashi, {\it {Multi-Natural Inflation}},  {\em Phys. Lett.
  B} {\bf 733} (2014) 241--246, [\href{http://arxiv.org/abs/1401.5212}{{\tt
  arXiv:1401.5212}}].

\bibitem{Czerny:2014xja}
M.~Czerny, T.~Higaki, and F.~Takahashi, {\it {Multi-Natural Inflation in
  Supergravity}},  {\em JHEP} {\bf 05} (2014) 144,
  [\href{http://arxiv.org/abs/1403.0410}{{\tt arXiv:1403.0410}}].

\bibitem{Higaki:2014pja}
T.~Higaki and F.~Takahashi, {\it {Natural and Multi-Natural Inflation in Axion
  Landscape}},  {\em JHEP} {\bf 07} (2014) 074,
  [\href{http://arxiv.org/abs/1404.6923}{{\tt arXiv:1404.6923}}].

\bibitem{Higaki:2014mwa}
T.~Higaki and F.~Takahashi, {\it {Axion Landscape and Natural Inflation}},
  {\em Phys. Lett. B} {\bf 744} (2015) 153--159,
  [\href{http://arxiv.org/abs/1409.8409}{{\tt arXiv:1409.8409}}].

\bibitem{Griest:1989wd}
K.~Griest and M.~Kamionkowski, {\it {Unitarity Limits on the Mass and Radius of
  Dark Matter Particles}},  {\em Phys. Rev. Lett.} {\bf 64} (1990) 615.

\bibitem{Parker:1969au}
L.~Parker, {\it {Quantized fields and particle creation in expanding universes.
  1.}},  {\em Phys. Rev.} {\bf 183} (1969) 1057--1068.

\bibitem{Parker:1971pt}
L.~Parker, {\it {Quantized fields and particle creation in expanding universes.
  2.}},  {\em Phys. Rev. D} {\bf 3} (1971) 346--356. [Erratum: Phys.Rev.D 3,
  2546--2546 (1971)].

\bibitem{Kuzmin:1998uv}
V.~Kuzmin and I.~Tkachev, {\it {Ultrahigh-energy cosmic rays, superheavy long
  living particles, and matter creation after inflation}},  {\em JETP Lett.}
  {\bf 68} (1998) 271--275, [\href{http://arxiv.org/abs/hep-ph/9802304}{{\tt
  hep-ph/9802304}}].

\bibitem{Chung:1998zb}
D.~J.~H. Chung, E.~W. Kolb, and A.~Riotto, {\it {Superheavy dark matter}},
  {\em Phys. Rev. D} {\bf 59} (1998) 023501,
  [\href{http://arxiv.org/abs/hep-ph/9802238}{{\tt hep-ph/9802238}}].

\bibitem{Chung:1998ua}
D.~J.~H. Chung, E.~W. Kolb, and A.~Riotto, {\it {Nonthermal supermassive dark
  matter}},  {\em Phys. Rev. Lett.} {\bf 81} (1998) 4048--4051,
  [\href{http://arxiv.org/abs/hep-ph/9805473}{{\tt hep-ph/9805473}}].

\bibitem{Kolb:1998ki}
E.~W. Kolb, D.~J.~H. Chung, and A.~Riotto, {\it {WIMPzillas!}},  {\em AIP Conf.
  Proc.} {\bf 484} (1999), no.~1 91--105,
  [\href{http://arxiv.org/abs/hep-ph/9810361}{{\tt hep-ph/9810361}}].

\bibitem{Chung:2001cb}
D.~J.~H. Chung, P.~Crotty, E.~W. Kolb, and A.~Riotto, {\it {On the
  Gravitational Production of Superheavy Dark Matter}},  {\em Phys. Rev. D}
  {\bf 64} (2001) 043503, [\href{http://arxiv.org/abs/hep-ph/0104100}{{\tt
  hep-ph/0104100}}].

\bibitem{Ling:2025nlw}
S.~Ling, A.~J. Long, E.~McDonough, and A.~Hayes, {\it {Superheavy Dark Matter
  from the String Theory Axiverse}},
  \href{http://arxiv.org/abs/2504.13256}{{\tt arXiv:2504.13256}}.

\bibitem{Kolb:2023ydq}
E.~W. Kolb and A.~J. Long, {\it {Cosmological gravitational particle production
  and its implications for cosmological relics}},  {\em Rev. Mod. Phys.} {\bf
  96} (2024), no.~4 045005, [\href{http://arxiv.org/abs/2312.09042}{{\tt
  arXiv:2312.09042}}].

\bibitem{Parker:2025jef}
L.~E. Parker, {\it {The Creation of Particles in an Expanding Universe}},
  \href{http://arxiv.org/abs/2507.05372}{{\tt arXiv:2507.05372}}.

\bibitem{Felder:1998vq}
G.~N. Felder, L.~Kofman, and A.~D. Linde, {\it {Instant preheating}},  {\em
  Phys. Rev. D} {\bf 59} (1999) 123523,
  [\href{http://arxiv.org/abs/hep-ph/9812289}{{\tt hep-ph/9812289}}].

\bibitem{Felder:1999pv}
G.~N. Felder, L.~Kofman, and A.~D. Linde, {\it {Inflation and preheating in NO
  models}},  {\em Phys. Rev. D} {\bf 60} (1999) 103505,
  [\href{http://arxiv.org/abs/hep-ph/9903350}{{\tt hep-ph/9903350}}].

\bibitem{Taya:2022gzp}
H.~Taya and Y.~Yamada, {\it {QFT approach to dressed particle processes in
  preheating and non-perturbative mechanism in kinematically-forbidden
  regime}},  {\em JHEP} {\bf 02} (2023) 048,
  [\href{http://arxiv.org/abs/2207.03831}{{\tt arXiv:2207.03831}}].

\bibitem{Dabrowski:2014ica}
R.~Dabrowski and G.~V. Dunne, {\it {Superadiabatic particle number in Schwinger
  and de Sitter particle production}},  {\em Phys. Rev. D} {\bf 90} (2014),
  no.~2 025021, [\href{http://arxiv.org/abs/1405.0302}{{\tt arXiv:1405.0302}}].

\bibitem{Dabrowski:2016tsx}
R.~Dabrowski and G.~V. Dunne, {\it {Time dependence of adiabatic particle
  number}},  {\em Phys. Rev. D} {\bf 94} (2016), no.~6 065005,
  [\href{http://arxiv.org/abs/1606.00902}{{\tt arXiv:1606.00902}}].

\bibitem{Yamada:2021kqw}
Y.~Yamada, {\it {Superadiabatic basis in cosmological particle production:
  application to preheating}},  {\em JCAP} {\bf 09} (2021) 009,
  [\href{http://arxiv.org/abs/2106.06111}{{\tt arXiv:2106.06111}}].

\bibitem{Planck:2018jri}
{\bf Planck} Collaboration, Y.~Akrami et~al., {\it {Planck 2018 results. X.
  Constraints on inflation}},  {\em Astron. Astrophys.} {\bf 641} (2020) A10,
  [\href{http://arxiv.org/abs/1807.06211}{{\tt arXiv:1807.06211}}].

\bibitem{ACT:2025fju}
{\bf ACT} Collaboration, T.~Louis et~al., {\it {The Atacama Cosmology
  Telescope: DR6 Power Spectra, Likelihoods and $\Lambda$CDM Parameters}},
  \href{http://arxiv.org/abs/2503.14452}{{\tt arXiv:2503.14452}}.

\bibitem{ACT:2025tim}
{\bf ACT} Collaboration, E.~Calabrese et~al., {\it {The Atacama Cosmology
  Telescope: DR6 Constraints on Extended Cosmological Models}},
  \href{http://arxiv.org/abs/2503.14454}{{\tt arXiv:2503.14454}}.

\bibitem{Ferreira:2025lrd}
E.~G.~M. Ferreira, E.~McDonough, L.~Balkenhol, R.~Kallosh, L.~Knox, and
  A.~Linde, {\it {The BAO-CMB Tension and Implications for Inflation}},
  \href{http://arxiv.org/abs/2507.12459}{{\tt arXiv:2507.12459}}.

\bibitem{SPT-3G:2025bzu}
{\bf SPT-3G} Collaboration, E.~Camphuis et~al., {\it {SPT-3G D1: CMB
  temperature and polarization power spectra and cosmology from 2019 and 2020
  observations of the SPT-3G Main field}},
  \href{http://arxiv.org/abs/2506.20707}{{\tt arXiv:2506.20707}}.

\bibitem{Kofman:2004yc}
L.~Kofman, A.~D. Linde, X.~Liu, A.~Maloney, L.~McAllister, and E.~Silverstein,
  {\it {Beauty is attractive: Moduli trapping at enhanced symmetry points}},
  {\em JHEP} {\bf 05} (2004) 030,
  [\href{http://arxiv.org/abs/hep-th/0403001}{{\tt hep-th/0403001}}].

\bibitem{Kikuchi:2023uqo}
S.~Kikuchi, T.~Kobayashi, K.~Nasu, and Y.~Yamada, {\it {Moduli trapping
  mechanism in modular flavor symmetric models}},  {\em JHEP} {\bf 08} (2023)
  081, [\href{http://arxiv.org/abs/2307.13230}{{\tt arXiv:2307.13230}}].

\bibitem{Dumlu:2010ua}
C.~K. Dumlu and G.~V. Dunne, {\it {The Stokes Phenomenon and Schwinger Vacuum
  Pair Production in Time-Dependent Laser Pulses}},  {\em Phys. Rev. Lett.}
  {\bf 104} (2010) 250402, [\href{http://arxiv.org/abs/1004.2509}{{\tt
  arXiv:1004.2509}}].

\bibitem{Li:2019ves}
L.~Li, T.~Nakama, C.~M. Sou, Y.~Wang, and S.~Zhou, {\it {Gravitational
  Production of Superheavy Dark Matter and Associated Cosmological
  Signatures}},  {\em JHEP} {\bf 07} (2019) 067,
  [\href{http://arxiv.org/abs/1903.08842}{{\tt arXiv:1903.08842}}].

\bibitem{Enomoto:2020xlf}
S.~Enomoto and T.~Matsuda, {\it {The exact WKB for cosmological particle
  production}},  {\em JHEP} {\bf 03} (2021) 090,
  [\href{http://arxiv.org/abs/2010.14835}{{\tt arXiv:2010.14835}}].

\bibitem{Taya:2020dco}
H.~Taya, T.~Fujimori, T.~Misumi, M.~Nitta, and N.~Sakai, {\it {Exact WKB
  analysis of the vacuum pair production by time-dependent electric fields}},
  {\em JHEP} {\bf 03} (2021) 082, [\href{http://arxiv.org/abs/2010.16080}{{\tt
  arXiv:2010.16080}}].

\bibitem{Hashiba:2021npn}
S.~Hashiba and Y.~Yamada, {\it {Stokes phenomenon and gravitational particle
  production \textemdash{} How to evaluate it in practice}},  {\em JCAP} {\bf
  05} (2021) 022, [\href{http://arxiv.org/abs/2101.07634}{{\tt
  arXiv:2101.07634}}].

\bibitem{Sou:2021juh}
C.~M. Sou, X.~Tong, and Y.~Wang, {\it {Chemical-potential-assisted particle
  production in FRW spacetimes}},  {\em JHEP} {\bf 06} (2021) 129,
  [\href{http://arxiv.org/abs/2104.08772}{{\tt arXiv:2104.08772}}].

\bibitem{Hashiba:2022bzi}
S.~Hashiba, S.~Ling, and A.~J. Long, {\it {An analytic evaluation of
  gravitational particle production of fermions via Stokes phenomenon}},  {\em
  JHEP} {\bf 09} (2022) 216, [\href{http://arxiv.org/abs/2206.14204}{{\tt
  arXiv:2206.14204}}].

\bibitem{Kofman:1997yn}
L.~Kofman, A.~D. Linde, and A.~A. Starobinsky, {\it {Towards the theory of
  reheating after inflation}},  {\em Phys. Rev. D} {\bf 56} (1997) 3258--3295,
  [\href{http://arxiv.org/abs/hep-ph/9704452}{{\tt hep-ph/9704452}}].

\bibitem{Peloso:2000hy}
M.~Peloso and L.~Sorbo, {\it {Preheating of massive fermions after inflation:
  Analytical results}},  {\em JHEP} {\bf 05} (2000) 016,
  [\href{http://arxiv.org/abs/hep-ph/0003045}{{\tt hep-ph/0003045}}].

\bibitem{Anchordoqui:2023etp}
L.~A. Anchordoqui and I.~Antoniadis, {\it {Large extra dimensions from
  higher-dimensional inflation}},  {\em Phys. Rev. D} {\bf 109} (2024), no.~10
  103508, [\href{http://arxiv.org/abs/2310.20282}{{\tt arXiv:2310.20282}}].

\bibitem{Antoniadis:2023sya}
I.~Antoniadis, J.~Cunat, and A.~Guillen, {\it {Cosmological perturbations from
  five-dimensional inflation}},  {\em JHEP} {\bf 05} (2024) 290,
  [\href{http://arxiv.org/abs/2311.17680}{{\tt arXiv:2311.17680}}].

\bibitem{Anchordoqui:2024amx}
L.~A. Anchordoqui and I.~Antoniadis, {\it {Primordial power spectrum of five
  dimensional uniform inflation}},  {\em Phys. Lett. B} {\bf 868} (2025)
  139673, [\href{http://arxiv.org/abs/2412.19213}{{\tt arXiv:2412.19213}}].

\bibitem{Hirose:2025pzm}
T.~Hirose, {\it {Analysis of inflationary models in higher-dimensional uniform
  inflation}},  {\em JHEP} {\bf 04} (2025) 077,
  [\href{http://arxiv.org/abs/2501.13581}{{\tt arXiv:2501.13581}}].

\bibitem{Cremades:2004wa}
D.~Cremades, L.~E. Ibanez, and F.~Marchesano, {\it {Computing Yukawa couplings
  from magnetized extra dimensions}},  {\em JHEP} {\bf 05} (2004) 079,
  [\href{http://arxiv.org/abs/hep-th/0404229}{{\tt hep-th/0404229}}].

\bibitem{Abe:2008sx}
H.~Abe, K.-S. Choi, T.~Kobayashi, and H.~Ohki, {\it {Three generation
  magnetized orbifold models}},  {\em Nucl. Phys. B} {\bf 814} (2009) 265--292,
  [\href{http://arxiv.org/abs/0812.3534}{{\tt arXiv:0812.3534}}].

\bibitem{Abe:2012ya}
H.~Abe, T.~Kobayashi, H.~Ohki, and K.~Sumita, {\it {Superfield description of
  10D SYM theory with magnetized extra dimensions}},  {\em Nucl. Phys. B} {\bf
  863} (2012) 1--18, [\href{http://arxiv.org/abs/1204.5327}{{\tt
  arXiv:1204.5327}}].

\bibitem{Abe:2012fj}
H.~Abe, T.~Kobayashi, H.~Ohki, A.~Oikawa, and K.~Sumita, {\it {Phenomenological
  aspects of 10D SYM theory with magnetized extra dimensions}},  {\em Nucl.
  Phys. B} {\bf 870} (2013) 30--54, [\href{http://arxiv.org/abs/1211.4317}{{\tt
  arXiv:1211.4317}}].

\bibitem{Linde:1982uu}
A.~D. Linde, {\it {Scalar Field Fluctuations in Expanding Universe and the New
  Inflationary Universe Scenario}},  {\em Phys. Lett. B} {\bf 116} (1982)
  335--339.

\bibitem{Starobinsky:1982ee}
A.~A. Starobinsky, {\it {Dynamics of Phase Transition in the New Inflationary
  Universe Scenario and Generation of Perturbations}},  {\em Phys. Lett. B}
  {\bf 117} (1982) 175--178.

\bibitem{Vilenkin:1982wt}
A.~Vilenkin and L.~H. Ford, {\it {Gravitational Effects upon Cosmological Phase
  Transitions}},  {\em Phys. Rev. D} {\bf 26} (1982) 1231.

\bibitem{Bergshoeff:2000zn}
E.~Bergshoeff, R.~Kallosh, and A.~Van~Proeyen, {\it {Supersymmetry in singular
  spaces}},  {\em JHEP} {\bf 10} (2000) 033,
  [\href{http://arxiv.org/abs/hep-th/0007044}{{\tt hep-th/0007044}}].

\bibitem{Bergshoeff:2001pv}
E.~Bergshoeff, R.~Kallosh, T.~Ortin, D.~Roest, and A.~Van~Proeyen, {\it {New
  formulations of D = 10 supersymmetry and D8 - O8 domain walls}},  {\em Class.
  Quant. Grav.} {\bf 18} (2001) 3359--3382,
  [\href{http://arxiv.org/abs/hep-th/0103233}{{\tt hep-th/0103233}}].

\bibitem{Fujita:2001bd}
T.~Fujita, T.~Kugo, and K.~Ohashi, {\it {Off-shell formulation of supergravity
  on orbifold}},  {\em Prog. Theor. Phys.} {\bf 106} (2001) 671--690,
  [\href{http://arxiv.org/abs/hep-th/0106051}{{\tt hep-th/0106051}}].

\bibitem{Kugo:2002js}
T.~Kugo and K.~Ohashi, {\it {Superconformal tensor calculus on orbifold in
  5D}},  {\em Prog. Theor. Phys.} {\bf 108} (2002) 203--228,
  [\href{http://arxiv.org/abs/hep-th/0203276}{{\tt hep-th/0203276}}].

\bibitem{Green:2009ds}
D.~Green, B.~Horn, L.~Senatore, and E.~Silverstein, {\it {Trapped Inflation}},
  {\em Phys. Rev. D} {\bf 80} (2009) 063533,
  [\href{http://arxiv.org/abs/0902.1006}{{\tt arXiv:0902.1006}}].

\bibitem{Flauger:2016idt}
R.~Flauger, M.~Mirbabayi, L.~Senatore, and E.~Silverstein, {\it {Productive
  Interactions: heavy particles and non-Gaussianity}},  {\em JCAP} {\bf 10}
  (2017) 058, [\href{http://arxiv.org/abs/1606.00513}{{\tt arXiv:1606.00513}}].

\bibitem{Wess:1992cp}
J.~Wess and J.~Bagger, {\em {Supersymmetry and supergravity}}.
\newblock Princeton University Press, Princeton, NJ, USA, 1992.

\bibitem{Dreiner:2008tw}
H.~K. Dreiner, H.~E. Haber, and S.~P. Martin, {\it {Two-component spinor
  techniques and Feynman rules for quantum field theory and supersymmetry}},
  {\em Phys. Rept.} {\bf 494} (2010) 1--196,
  [\href{http://arxiv.org/abs/0812.1594}{{\tt arXiv:0812.1594}}].

\bibitem{Elizalde:1995hck}
E.~Elizalde, {\em {Ten physical applications of spectral zeta functions}},
  vol.~35.
\newblock 1995.

\bibitem{Arkani-Hamed:1999lsd}
N.~Arkani-Hamed, L.~J. Hall, D.~Tucker-Smith, and N.~Weiner, {\it {Solving the
  hierarchy problem with exponentially large dimensions}},  {\em Phys. Rev. D}
  {\bf 62} (2000) 105002, [\href{http://arxiv.org/abs/hep-ph/9912453}{{\tt
  hep-ph/9912453}}].

\end{thebibliography}\endgroup
\end{document}